\documentclass[11pt]{article}
\usepackage{amsmath}
\usepackage{graphicx}
\usepackage{enumerate}
\usepackage{natbib}
\usepackage{url} 
\usepackage{amsmath}
\usepackage{makecell}
\usepackage{subcaption}

\newcommand{\prob}[1]
  {\text{P}\left\{#1\right\}}

\newcommand{\probf}[2]
  {\text{P}_{#1}\left\{#2\right\}}
  
\newcommand{\mb}[1]
   {\boldsymbol{#1}}
   
\newcommand{\Ef}[2]
  {\mbox{E}_{#1}\left\{#2\right\}}
   
\newcommand{\probh}[1]
  {\hat{\text{P}}\left\{#1\right\}}

\newcommand{\var}[1]
  {\mbox{Var}\left\{#1\right\}}
  
\newcommand{\Cov}[1]
  {\mbox{Cov}\left\{#1\right\}}
  
\newcommand{\E}[1]
  {\mbox{E}\left\{#1\right\}}
  
\newcommand{\cov}[1]
  {\mbox{Cov}\left\{#1\right\}}

\newcommand{\convDistr}[0]
  {\stackrel{d}{\longrightarrow}}

\newcommand{\convProb}[0]
  {\stackrel{p}{\longrightarrow}}

\newcommand{\indicat}[1]
  {\mbox{I}\left\{#1\right\}}

\newcommand{\blind}{1}

\begin{document}

\def\spacingset#1{\renewcommand{\baselinestretch}%
{#1}\small\normalsize} \spacingset{1}


\if1\blind
{
  \title{\bf The Probability of Improved Prediction: a new concept in statistical inference}
  \author{Olivier Thas\thanks{
    The authors gratefully acknowledge Research Foundation - Flanders (FWO) for providing the funding for this research (Grant nr. G0D1221N)}\hspace{.2cm}\\
    Interuniversity Institute for Biostatistics \& statistical Bioinformatics\\ (I-BioStat), Hasselt University, 3590 Diepenbeek, Belgium\\
    Department of Applied Mathematics, Computer Science and Statistics,\\ Faculty of Sciences,Ghent University, Krijgslaan 281, 9000 Ghent, Belgium\\
    National Institute of Applied Statistics Research Australia (NIASRA),\\ University of Wollongong, Northfield Ave, Wollongong, NSW 2522, Australia\\
    and \\
    Stijn Jaspers \\
    Interuniversity Institute for Biostatistics \& statistical Bioinformatics\\ (I-BioStat), Hasselt University, 3590 Diepenbeek, Belgium}
  \maketitle
} \fi

\if0\blind
{
  \bigskip
  \bigskip
  \bigskip
  \begin{center}
    {\LARGE\bf The Probability of Improved Prediction: a new concept in statistical inference}
\end{center}
  \medskip
} \fi

\bigskip
\begin{abstract}
In an attempt to provide an answer to the increasing criticism against $p$-values and to bridge the gap between statistical inference and prediction modelling, we introduce the probability of improved prediction (PIP). In general, the PIP is a probabilistic measure for comparing two competing models. Three versions of the PIP and several estimators are introduced and the relationships between them, $p$-values and the mean squared error are investigated. The performance of the estimators is assessed in a simulation study. An application shows how the PIP can support $p$-values to strengthen the conclusions or possibly point at issues with e.g. replicability. 
\end{abstract}

\noindent%
\vfill

\newpage
\spacingset{1.4} 
\section{Introduction}
\label{sec:intro}

Hypothesis testing is one of the major concepts in statistical inference and it is still a well accepted paradigm as part of the Scientific Method in many empirical sciences. In particular, the Scientific Method advises to first formulate a research hypothesis that is subsequently translated into a null and alternative hypothesis. After the data collection, a statistical test is applied for testing the null against the alternative hypothesis, and in most scientific disciplines the convention is to compare the $p$-value with a nominal threshold of 5\%. If $p<5\%$, then the null hypothesis is rejected in favor of the alternative, which often corresponds to a {\em positive} research result, and if $p>5\%$, then the null cannot be rejected and the study ends inconclusively. 

Despite the firm theoretical basis of hypothesis testing, (1) its application, (2) how researchers use $p$-values in their communication, and (3) the publication policy of many journals, have caused several biases in research. We name a few. First, the {\em publication bias} refers to the observation that mostly positive results can be published in scientific journals, and consequently many, still informative results are hidden from the community. A second concern is related to the use of hypothesis testing: researchers might test several hypotheses, until one gives a ``significant'' result, which is subsequently published as a positive result without reporting the $p$-fishing expeditions that resulted in this finding. This process is known to give an increased, uncontrolled false positive rate. Finally, some researchers still translate $p>5\%$ into a proof of a negative finding, resulting in the publication of false negative results. 

The criticism against the conventional use of $p$-values is sounding louder and louder these days. In 2019 an issue of The American Statistician was completely devoted to this topic. Also the journal Nature has published several comments on the (mis)use of $p$-values in research \citep[e.g.][]{Baker2016, Nuzzo2014, Amrhein2019}. A paper by \cite{Amrhein2019} is supplemented with 859 signatures of scientists who support the recommendation to abandon the current conventional use of $p$-values. These recent papers have provoked many reactions and even journals have altered their publication guidelines, or make their readers aware of potential pitfalls \citep[e.g.][]{Drummond2020,Dunkler2020,Harrington2019, Ioannidis2019}.

On a personal note, while teaching my basic statistics course, I (the first author) explain the students the role of statistical inference in the Scientific Method, but each year I feel uncomfortable when I hear myself telling the students that science aims at improving models for describing the world around us. These models should predict observations, whereas many statistical methods target parameters that have an interpretation at the level of the population average. Although there is often a relation between improved predictability by considering an additional predictor in a statistical model, and the size of the parameter corresponding to this predictor, hypothesis testing does not directly address the improved predictability.

\cite{Wasserstein2019} summarises the contributions and suggestions made by the authors of all 43 papers in the 2019 special issue of The American Statistician. Many authors make suggestions in line with: (1) no longer simply apply the $p<5\%$ rule; (2) report the $p$-value up to a relevant precision; (3) no need to abandon the $p$-value, but it should be complemented with other relevant statistics (e.g. point estimates, confidence intervals, ...); (4) statistical results should be interpreted by researchers with good substantial knowledge (i.e. no longer computing statistics and blindly applying simple binary decision rules). In this light, the new inferential concept that we propose in this paper must be considered as an addition to the existing conventional tools, rather than a replacement. Our concept aims at bringing statistical inference and prediction model selection closer together, as we think this connection is implicitly made in the Scientific Method.

Some proposals have been made for new statistics that may complement the traditional inferential tools. For example: (1) compute $p$-values evaluated under a meaningful alternative \citep{Amrhein2019,Greenland2019} or under a range of null hypotheses (composite null hypothesis, second generation $p$-values) \citep{Blume2019}; (2) augment $p$-values with information about the false positive risk \citep{Colquhoun2019}. Only few suggestions make a connection between statistical inference and prediction models. In this sense, \cite{Billheimer2019} comes closest to our method. He suggests that inference should be based on the predictive distribution (PD), which is the distribution of the difference between the prediction, which is based on a model fitted on (training) data, and an independent outcome that has to be predicted. He stresses the importance of the variance of the PD. 

Formal definitions of the new concepts and their properties are given in subsequent sections, but as a matter of introduction, we start here with a description of the setting and the overall rationale. Consider the setting in which we have two nested models for an outcome variable and one model is based on one additional regressor, covariate or predictor, whatever terminology is preferred. With $Y$ the notation for the outcome and $\mb{x}$ the notation for a vector of predictors, statistical models are often of the form $\E{Y\mid \mb{x}} = m(\mb\beta,\mb{x})$, with $\mb\beta$ a parameter vector and $m(\cdot)$ representing the {\em model}. However, in the context of prediction modelling, there is often no focus on the conditional mean outcome $\E{Y\mid \mb{x}}$ and so $m(\mb\beta,\mb{x})$ can also refer to a prediction model aiming at predicting an outcome. We consider two models, $m^0(\mb\beta^0,\mb{x})$ and $m^1(\mb\beta^1,\mb{x})$, with two different parameter vectors $\mb\beta^0$ and $\mb\beta^1$. In general, the two models may depend on the same vector of predictors, $\mb{x}$, but for now we limit the discussion to the setting in which $m^0$ has one predictor less than $m^1$ such that model $m^0$ is nested within model $m^1$; let $x_1$ denote this predictor. 
If these are additive linear models for the conditional mean of the outcome, then we could consider testing the hypothesis that the $\beta^1$ parameter corresponding to $x_1$ equals zero. However, if prediction is the final objective, then both models could be compared to one another in terms of e.g. the mean squared error. These assessments are typically made based on a data set, which can be referred to as the sample data or the training data, depending on the context. We will use the notation ${\cal{O}}$ for this data set, which contains $n$ observed outcomes and $d$-dimensional predictors. Prediction is always about predicting new, yet unseen outcomes. Let $(\mb{X}^*,Y^*)$ denote a vector of a new predictor vector $\mb{X}^*$ and a new outcome $Y^*$, which are not part of the observed data ${\cal{O}}$ and for which $F^*$ denotes their joint distribution, which may depend on parameters, denoted by $\mb\nu$. Note that we do not assume that any of the two models $m^0$ and $m^1$ agrees with $F^*$. However, if e.g. $m^1$ agrees with $F^*$ (i.e. it has the conditional mean $\E{Y^*\mid \mb{X}^*}=m^1(\mb{X}^*)$), then the $\mb\beta^1$ parameter is part of $\mb\nu$. The parameter estimates of $\mb\beta^0$ and $\mb\beta^1$ are both calculated from the data set ${\cal{O}}$. These estimates are denoted by $\hat{\mb\beta}^0({\cal{O}})$ and $\hat{\mb\beta}^1({\cal{O}})$, or by their shorthand notation $\hat{\mb\beta}^0$ and $\hat{\mb\beta}^1$.
With this notation, the mean squared error of a model can be formulated as
\begin{equation}
  \label{Eq_MSE}
  \text{MSE} = \Ef{X^*,Y^*}{\left(Y^* - m(\hat{\mb\beta};\mb{X}^*)\right)^2 \mid {\cal{O}}}.
\end{equation}
\cite{Hastie2009} refer to this definition as the {\em generalisation error} and argue that this is the most relevant version of the MSE for the evaluation of a prediction model: it only assesses the performance for predicting new, unseen outcomes, of the single prediction model that will be used in practice, that is the model fitted (or trained) on the observed data ${\cal{O}}$. Since the MSE depends on the unknown distribution of $(\mb{X}^*,Y^*)$, its estimation is less straightforward. 

Also the $p$-value can be formulated with this notation. This time we consider a sample of $n$ i.i.d. unseen observations $(\mb{X}^*,Y^*)$, but under the restriction that the null hypothesis holds. This random sample is denoted by ${\cal{O}}^*$. With $T({\cal{O}})$ the test statistic as a function of sample data and for which large values point into the direction of the alternative hypothesis, the $p$-value can be written as
\begin{equation}
  \label{Eq_PValue}
  p = \probf{{\cal{O}}^*}{T({\cal{O}}^*) \geq T({\cal{O}}) \mid {\cal{O}}}.
\end{equation}
So far, the MSE and the $p$-value show great similarities in terms of what is considered fixed and random. However, there are a few fundamental differences. First, the motivation for the use of the $p$-value comes from its distribution over repeated sample data sets ${\cal{O}}$ under the restriction of the null hypothesis (i.e. ${\cal{O}} \stackrel{d}{=} {\cal{O}}^*$). This gives a uniform distribution of the $p$-value and its control of the type I error rate upon using a hard threshold. A second difference comes from the number of models involved in their definitions. The MSE refers to only a single model. When models have to be compared, e.g. two nested models $m^0$ and $m^1$, the MSE will be computed for the two models and their MSEs will be compared. The $p$-value, on the other hand, compares two nested models. Finally, the MSE as presented in Equation (\ref{Eq_MSE}) must first be estimated before it can be used. This typically happens via cross-validation or bootstrap resampling schemes. In this way the sample-based MSE is seen as an estimator of the generalisation error of Equation (\ref{Eq_MSE}), and it is often reported with a standard error. The latter may sometimes be used in the decision process to decide which of two (or more) models is the best; however, such decision rules are not based on a firm theory.  

We propose a new concept: the probability of improved prediction (PIP), which, in our current context, may take the form
\begin{equation}
  \label{Eq_PIPCond_Intro}
  \probf{X^*,Y^*}{\left(Y^* - m^1(\hat{\mb\beta}^1;\mb{X}^*)\right)^2 < \left(Y^* - m^0(\hat{\mb\beta}^0;\mb{X}^*)\right)^2 \mid {\cal{O}}}.
\end{equation}
This can be interpreted as follows. For a given dataset ${\cal{O}}$, used for fitting (or training) the two models $m^0$ and $m^1$, it expresses the relative frequency of instances that model 1 gives a better prediction than model 0. This frequency, or probability, is defined over the distribution of future, unseen observations and the interpretation of ``better'' is in terms of the squared loss. Just like the $p$-value and the MSE, it is defined conditional on the observed data, so that conclusions based on the PIP directly relate to the models to be used for later purposes such as prediction. We therefore will refer to the PIP of Equation (\ref{Eq_PIPCond_Intro}) as the {\em conditional PIP}. However, just like the generalisation error, the PIP first needs to be estimated, and details will follow later in this paper. In this sense, the PIP and MSE have much in common, but they address other aspects: the MSE is about how well an individual model performs in predicting outcomes, whereas the PIP compares two models and quantifies how often one model does better then the other. However, from the perspective of frequentist statistical inference, we may be interested in the effect of a predictor, on average over repeated samples. For this purpose, we also introduce the {\em expected PIP}, 
\begin{equation*}
  \label{Eq_PIPExp_Intro}
  \probf{X^*,Y^*,{\cal{O}}}{\left(Y^* - m^1(\hat{\mb\beta}^1;\mb{X}^*)\right)^2 < \left(Y^* - m^0(\hat{\mb\beta}^0;\mb{X}^*)\right)^2},
\end{equation*}
which is the expectation of the conditional PIP over the distribution of the sample data. Just like the conditional PIP and the MSE, it needs to be estimated. 

In the remainder of this paper, we will first more generally define the different versions of the PIP, as well as methods for their estimation (Section 2). To make the connection with classical hypothesis testing more explicit, we will work out some more details for the PIP in the very simple case of two nested normal linear regression models, with a 0/1 dummy variable (the hypothesis test is thus the two-sample $t$-test). At the end of Section 2, we will discuss nonparametric estimation methods for the PIP, which allows for the use of almost arbitrary complicated models $m^0$ and $m^1$ and model fitting (or training) methods, making the PIP also applicable to machine learning models and methods.  
In Section 3 results of a simulation study are presented; this empirical study focuses on the properties of the various estimators of the PIP. An illustration of the PIP on several datasets is given in Section 4, and we conclude in Section 5 with a further discussion on this new concept.

\section{The Probability of Improved Prediction and its Estimators}
\label{sec:meth}
In this section, we first introduce three types of PIPs. Next we propose a few estimators. In a concluding subsection, these general ideas are applied in the simple setting of the two-sample t-test.

\subsection{Three Types of PIPs}

Upon using the notation and terminology introduced in Section \ref{sec:intro}, we start with the definition of the {\em theoretical PIP} in terms of a general loss function. Let $m^0(\mb\beta^0,\mb{X}^*)$ and $m^1(\mb\beta^1,\mb{X}^*)$ denote two prediction models for predicting an outcome $Y^*$ for a given predictor vector $\mb{X}^*$. The two prediction models may depend on two different parameter vectors, $\mb\beta^0$ and $\mb\beta^1$. Let $L(m(\mb{X}^*),Y^*)$ denote a loss function evaluated in a prediction model $m$ for predicting $Y^*$. The squared loss that we used in the introduction is an obvious choice for continuous outcomes. 
The theoretical PIP is then defined as 
\begin{equation}
     p_\text{th}(\mb\beta^0,\mb\beta^1,F^*,\mb{\nu}):=\probf{X^*,Y^*}{L(m^1(\mb\beta^1,\mb{X^*}),Y^*) < L(m^0(\mb\beta^0,\mb{X^*}),Y^*)}
\label{eq:theor_pip}
\end{equation}
which quantifies how more often model $m^1$ gives better predictions than model $m^0$. It should be noted that the PIP does not depend on the training data, nor on the distribution of the training data. It does depend on two models and their parameter vectors, and the probability is defined over the distribution of $(\mb{X}^*,Y^*)$, of which the distribution function is denoted by $F^*(\mb{\nu})$ with parameter vector $\mb\nu$. As shown in Appendix \ref{A_theoretical}, for the squared error loss function, the theoretical PIP can be expressed as
\begin{equation}
  \label{pip_th}
  \int_{\cal{X^*}} \probf{Y^*|X^*}{m^1(\mb\beta^1)^2- m^0(\mb\beta^0)^2 < 2Y^*(m^1(\mb\beta^1)-m^0(\mb\beta^0))|X^*=x^*}dF^*_{X^*}(x^*),
\end{equation}
with ${\cal{X^*}}$ the sampling space of $X^*$. A further simplification of this equation needs careful consideration of the inequality, for it depends on the sign of $m^1(\mb\beta^1)-m^0(\mb\beta^0)$, which is determined by both the model parameters ($\mb\beta^0$ and $\mb\beta^1$) and $x^*$. Section \ref{Example_TwoSample} and Appendix \ref{A_theoretical} show some examples for the case of two linear regression models. 
The theoretical PIP may not be of immediate practical use as it only compares two completely specified models, but it may be of interest to estimate this PIP (see later). 


In terms of a general loss function $L(\cdot,\cdot)$, the {\em conditional PIP} is defined as
\begin{equation}
  \label{Eq_condPIP}
  p_\text{cond}(\hat{\mb\beta}^0,\hat{\mb\beta}^1,F^*,{\mb\nu}):=\probf{X^*,Y^*}{L(m^1(\hat{\mb\beta}^1,\mb{X^*}),Y^*) < L(m^0(\hat{\mb\beta}^0,\mb{X^*}),Y^*) \mid {\cal{O}}}.
\end{equation}
This probability still only refers to the distribution of $(\mb{X}^*,Y^*)$, and the explicit conditioning on ${\cal{O}}$ is used to indicate that the sampling distributions of the parameter estimators $\hat{\mb\beta}^0$ and $\hat{\mb\beta}^1$ are not considered. Note that this expression resembles the structure of the MSE (Equation (\ref{Eq_MSE})) and the $p$-value (Equation (\ref{Eq_PValue})), so it may also be an informative instrument in statistical inference or prediction modelling. Just like the MSE and the $p$-value, the conditional PIP summarises a property of a model (or models) that is relevant for future, unseen observations $(\mb{X}^*,Y^*)$, based on the information provided by the observed data ${\cal{O}}$. 

However, just like the theoretical PIP, the conditional PIP cannot be computed, because in practice the distribution $F^*$ of $(\mb{X}^*,Y^*)$ is unknown. Estimators will be discussed later. 

Sometimes one may not be interested in the information within a single sample ${\cal{O}}$, but rather in the performance of the model building or parameter estimation procedures. For this purpose the {\em expected PIP} may be appropriate. It is defined as the expectation of the conditional PIP over the distribution of the sample data ${\cal{O}}$,
\begin{eqnarray*}
 p_\text{exp}(\mb\beta^0,\mb\beta^1,F^*,F_{\cal{O}},\mb\nu) 
 &:=& \Ef{{\cal{O}}}{\probf{X^*,Y^*}{L(m^1(\hat{\mb\beta}^1,\mb{X^*}),Y^*) < L(m^0(\hat{\mb\beta}^0,\mb{X^*}),Y^*) \mid {\cal{O}}}} \\
 &=& \probf{X^*,Y^*,{\cal{O}}}{L(m^1(\hat{\mb\beta}^1,\mb{X^*}),Y^*) < L(m^0(\hat{\mb\beta}^0,\mb{X^*}),Y^*)}.
\end{eqnarray*}

\subsection{Plug-in Estimators}

We will discuss two types of estimators: plug-in and nonparametric (re)sampling-based estimators. The former will be discussed here, and the latter in the next section.

According to the general plug-in principle, all unknown parameters are replaced by their sample estimators. Table \ref{Tab_WhatIsEstimated} shows what parameters are unknown in each of the three PIPs. Note that the $\mb\beta^j$ ($j=0,1$) parameters are indicated as known for the conditional PIP. This is because the conditional PIP is defined conditional on the single observed sample data set ${\cal{O}}$ from which these parameters are estimated. The presence of fixed estimates is a fundamental part of the definition of this PIP, and hence the estimators are not to be considered as unknown. For the other two PIPs, these $\beta$-parameters are unknown and can be replaced with their parameter estimates. 

Apart from the $\beta$-parameters, all three PIPs depend on $F^*$, which is the unknown distribution function of the future, unseen observations $(\mb{X}^*,Y^*)$. We consider two approaches: (1) a parametric method that involves distributional assumptions, and (2) nonparametric methods that do not impose strong distributional assumptions. In this section we proceed with the former. In particular, we will assume that $F^*$ agrees with $m^1$, which is by construction a more flexible model than $m^0$ (e.g. $m^1$ embeds $m^0$). In the examples provided later, $m^1$ is a model for the conditional mean outcome $\E{Y\mid \mb{X}}$, and hence further distributional assumptions are required for the complete specification of $F^*$. For example, we may want to assume that $Y \mid \mb{X} \sim N(m^1(\mb\beta^1,\mb{X}),\sigma_1^2)$. In this construction, the estimation of $F^*$ boils down to estimating $\mb\beta^1$ and $\sigma_1^2$. Note that $\sigma_1^2$ and $\mb\beta^1$ are here part of the $\mb\nu$ parameter in our general setup. 

Hence, upon using the parametric plug-in strategy outlined in the previous section, we get the plug-in estimator of the theoretical PIP as $p_\text{th}(\hat{\mb\beta}^0,\hat{\mb\beta}^1,F^*,\hat\nu)$, with $F^*$ a normal distribution. In practice, Equation (\ref{pip_th}) can be used when the squared loss is considered. 
The plug-in estimate of the conditional PIP can be obtained in a similar fashion, but note that this PIP is by definition already a function of $\hat{\mb\beta}^0$ and $\hat{\mb\beta}^1$. Hence, only $F^*$ and $\mb\nu$ have to be specified or estimated. Under the same model assumptions as we made for the theoretical PIP, we replace $F^*$ by a normal distribution and set $\hat{\mb\nu}^t=(\hat{\mb\beta}^{1t},\hat\sigma_1^2)$.
It is interesting to note that the plug-in estimates of the theoretical and conditional PIP coincide! However, they are conceptually different. In the plug-in estimator of the theoretical PIP, all appearances of $\hat{\mb\beta}^0$ and $\hat{\mb\beta}^1$ are considered as random variables, whereas for the conditional PIP, only the appearance of $\hat{\mb\beta}^1$ as part of $\hat{\mb\nu}$ is considered as a random variable. 

Finally, for the estimation of the expected PIP, we see from its definition (and Table \ref{Tab_WhatIsEstimated}) that also $F_{\cal{O}}$, the distribution of the sample observations, is unknown and needs to be estimated. It is natural to assume that the sample observations in ${\cal{O}}$ have the same distribution as the future, unseen observations, and hence, under the additional assumption that all sample observations are i.i.d, we set $F_{\cal{O}}$ to the product of $n$ times $F^*$. The plug-in estimator of the expected PIP is therefore given by $p_\text{exp}(\hat{\mb\beta}^0,\hat{\mb\beta}^1,F^*,F_{\cal{O}},\hat{\mb\nu})$, with $F^*$ and $F_{\cal{O}}$ normal distributions and with the parameter estimators as before. In Section \ref{Example_TwoSample} more explicit formulae are given for a simple example. 

In this section the construction of plug-in estimators is restricted to a scenario in which $m^1$ embeds the model $m^0$, and limited to normal distributions. Obviously, other distributions for $F^*$ and $F_{\cal{O}}$ are also possible. In Section \ref{Nonpara} we propose nonparametric PIP estimators for which distributional assumptions are no longer required. 

\begin{table}[htbp!]
\caption{For each of the three types of PIP, the letter U indicates that the corresponding parameter is unknown. An empty entry indicates that the corresponding parameter is either known or not part of the definition of the PIP. 
\label{Tab_WhatIsEstimated}}
  
\begin{center}
    \begin{tabular}{lccc}
    \hline
     & theor. PIP & cond. PIP & exp. PIP \\
    \hline
$\mb\beta^0, \mb\beta^1$ & U &   & U \\
$F^*$                    & U & U & U\\
$\mb\nu$                 & U & U & U \\
$F_{\cal{O}}$            &   &   & U \\
    \hline
    \end{tabular}
\end{center}
\end{table}

\subsection{Nonparametric estimators}
\label{Nonpara}

The distribution $F^*$ is generally not known in practice and assuming that its mean corresponds to $m^1$ might be incorrect. For this reason, several nonparametric estimators of the PIPs are considered. In particular, we will look at sample splitting and cross-validation. Both approaches start by dividing the observed data set ${\cal{O}}$ into a training and a test set. Next, the training set is used for calculating the parameter estimates of the two models under consideration, while the test set is used to nonparametrically estimate the PIP by counting how many times model $m^1$ provides a better prediction than model $m^0$ in terms of the loss function $L$. The difference between both approaches lies in the way the actual splitting into a training and test set is conducted. 

For sample splitting, the data set is randomly split into two halves. Half of the observations are considered as a training set that is used for fitting or training the two models, while the other half is used as a test set for estimating the PIP. Hence, the parameters are estimated once and also only a single estimate for the PIP is calculated. 
In particular, with ${\cal{O}}^\text{Tr}$ and ${\cal{O}}^\text{Te}$ denoting the training and test data sets, and with $\hat{\mb\beta}^j({\cal{O}})$ denoting an estimate of $\mb\beta^j$ based on the data set ${\cal{O}}$, the split-sample PIP estimator is given by
\begin{equation}
    \label{Eq_SS_PIP}
    \frac{1}{\Vert {\cal{O}}^\text{Te}\Vert} \sum_{(\mb{x}^*,y^*) \in {\cal{O}}^\text{Te}} \indicat{L(m^1(\hat{\mb\beta}^1({\cal{O}}^\text{Tr}),\mb{x}^*),y^*) < L(m^0(\hat{\mb\beta}^0({\cal{O}}^\text{Tr}),\mb{x}^*),y^*)}.
\end{equation}

On the other hand, with $k$-fold cross-validation, the observed data are randomly divided into $k$ folds, and each fold is in turn considered to be the test set, while the remaining $k-1$ folds are used as the training set. As such, $k$ estimates of the PIP are obtained and their average forms the final PIP estimate. We will focus on leave-one-out cross-validation ($k=n$) and (repeated) 5-fold cross-validation (k=5). 
With ${\cal{O}}^\text{Tr}_j$ and ${\cal{O}}^\text{Te}_j$ the training and test data set for the $j$th fold ($j=1,\ldots, k$), the $k$-fold cross-validation PIP estimator is given by
\begin{equation*}
    \frac{1}{k} \sum_{j=1}^k 
    \frac{1}{\Vert {\cal{O}}_j^\text{Te}\Vert} \sum_{(\mb{x}^*,y^*) \in {\cal{O}}_j^\text{Te}} \indicat{L(m^1(\hat{\mb\beta}^1({\cal{O}}_j^\text{Tr}),\mb{x}^*),y^*) < L(m^0(\hat{\mb\beta}^0({\cal{O}}_j^\text{Tr}),\mb{x}^*),y^*)}.
\end{equation*}
In case of {\em repeated} $k$-fold cross-validation, the random division into $k$ folds is repeated $M$ times, leading to ${\cal{O}}^\text{Tr,m}_j$ and ${\cal{O}}^\text{Te,m}_j, m=1,\ldots,M$. The PIP estimator is then given as the average of the $M$ $k$-fold cross-validation estimators:
\begin{equation*}
    \frac{1}{M}\sum_{m=1}^M  \frac{1}{k} \sum_{j=1}^k 
    \frac{1}{\Vert {\cal{O}}_j^\text{Te,m}\Vert} \sum_{(\mb{x}^*,y^*) \in {\cal{O}}_j^\text{Te,m}} \indicat{L(m^1(\hat{\mb\beta}^1({\cal{O}}_j^\text{Tr,m}),\mb{x}^*),y^*) < L(m^0(\hat{\mb\beta}^0({\cal{O}}_j^\text{Tr,m}),\mb{x}^*),y^*)}.
\end{equation*}
A lower $(1-\alpha)100\%$ confidence bound for the repeated $k$-fold CV PIP, determined as the  $\alpha^{th}$ quantile of the $M$ $k$-fold cross-validation estimators, will later be used.  

\subsection{Some Relationships between PIPs}
\label{S_RelationsBetweenPIPs}

\begin{figure}[ht!]
    \centering
    \includegraphics[scale=0.25]{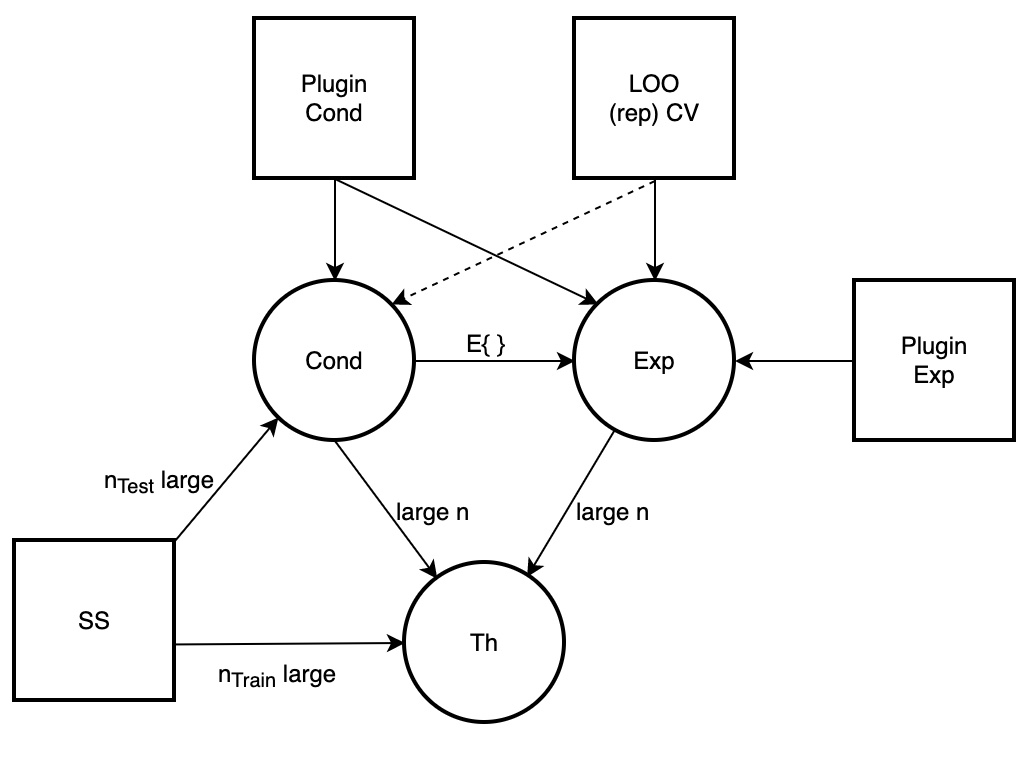}
    \caption{A schematic overview of the relationships between the three types of PIP (circles) and the estimators (rectangles). Lines between the PIPs indicate how one PIP depends on the other, and a line between an estimator and a PIP indicates that the estimator is designed to estimate that PIP. Although the resampling estimators are supposed to estimate the expected PIP, we will also evaluate them as estimators of the conditional PIP (shown as a dashed line). The sample size is denoted by $n$, and the sample sizes of the training and test datasets are denoted by $n_\text{Train}$ and $n_\text{Test}$, respectively.}
    \label{fig_relationships}
\end{figure}

The expected PIP is obviously related to the conditional PIP, for the former is the expectation of the latter. 
However, the expected and conditional PIPs are also related to the theoretical PIP. In particular, under mild regularity conditions on the loss function $L$, strong convergence of the estimators $\hat{\mb\beta}^0$ and $\hat{\mb\beta}^1$ implies that the conditional PIP, as $n\rightarrow \infty$, converges in probability to the theoretical PIP $p_\text{th}(\tilde{\mb{\beta}}^0,\tilde{\mb{\beta}}^1,F^*,F_{\cal{O}},\mb\nu)$, with $\tilde{\mb{\beta}}^0$ and $\tilde{\mb{\beta}}^1$ the probability limits of their estimators. Hence, also the expected PIP approaches the theoretical PIP with increasing sample size. 
These relationships are schematically presented in Figure \ref{fig_relationships}. The figure also shows what estimator is designed to estimate what PIP. The plug-in estimator/PIP combinations have been discussed in the earlier sections, and the following paragraphs discuss the resampling estimators as estimators of the expected, conditional and/or theoretical PIP. 

In the split-sample estimator of Equation (\ref{Eq_SS_PIP}), two sample sizes are involved: $n_\text{Train}=\#{\cal{O}}^\text{Tr}$ for the training data, and $n_\text{Test}=\#{\cal{O}}^\text{Te}$ for the test data. We consider two scenarios. 
First, when $n_\text{Test}$, $x^*$ and $y^*$ are fixed, and under mild regularity conditions, as $n_\text{Train}\rightarrow \infty$, $
  L(m^1(\hat{\mb\beta}^j({\cal{O}}^\text{Tr}),\mb{x}^*),y^*) \convProb L(m^1(\tilde{\mb\beta}^j,\mb{x}^*),y^*)$, with $\tilde{\mb\beta}^j$ the probability limit of $\hat{\mb\beta}^j({\cal{O}}^\text{Tr})$ ($j=0,1$). In this case, the split-sample estimator is asymptotically unbiased for the theoretical PIP (evaluated in the probability limits of the parameter estimators).

Second, when $n_\text{Train}$ and ${\cal{O}}^\text{Tr}$ are fixed, and under mild regularity conditions, as $n_\text{Test}\rightarrow \infty$, the split-sample estimator converges in probability to 
\[
  \probf{X^*,Y^*}{L(m^1(\hat{\mb\beta}^1({\cal{O}}^\text{Tr}),\mb{X^*}),Y^*) < L(m^0(\hat{\mb\beta}^0({\cal{O}}^\text{Tr}),\mb{X^*}),Y^*) \mid {\cal{O}}^\text{Tr}},
\] 
which is the conditional PIP for the (fixed) sample ${\cal{O}}^\text{Tr}$. 

In practice, when both the training and the test datasets are large, the split-sample estimator may be considered as an estimator of both the conditional and theoretical PIP. For small sample sizes, it is less clear. This will be further empirically investigated in the simulation study of Section \ref{sec:sim}.

The (repeated) $k$-fold CV estimator can be see as an average of conditional PIPs, i.e. for each fold a conditional PIP is computed and the final estimator is an average of these conditional PIPs over multiple independent folds. In this way, the CV estimator can be seen as an estimator of the expected PIP. This observation is similar to the conclusion made in \cite{Bates2021}, where the CV MSE was found to provide better estimates for the expected average prediction error (expectation over $\mb{X}$ and $Y$), rather than for the average prediction error conditional on the observed data $(\mb{X},Y)$.    

\subsection{The Two-Sample Setting}
\label{Example_TwoSample}

As a very simple example, we consider the two sample setting presented as two nested linear models with normal error terms. In particular,  $m^0(\mb\beta^0,x) = \beta_0^0$ and $m^1(\mb\beta^1,x) = \beta_0^1 + \beta_1^1 x$ with $x$ a $0/1$ dummy variable. The reason for choosing for this very simple scenario is that (1) explicit formulae can be easily developed, and (2) it allows us to make the link with hypothesis testing explicit without losing focus because of complicated models. These results directly translate to the more general case of two nested normal linear regression models, for which details are provided in Appendix \ref{A_theoretical}. More complicated models, for which hypothesis testing is no longer straightforward, are discussed in Section \ref{Nonlinear}.

In addition to the specification of the two nested models $m^0$ and $m^1$, we assume that the sample observations are distributed according to the larger model, meaning that  $Y_i \mid x_i \sim N(m^1(\mb\beta^1,x_i),\sigma_1^2)$, $i=1,\ldots, n$. This model specifies both $F^*$ and $F_{\cal{O}}$. 
It is further assumed that the regressor $X$ is a $0/1$ random binary group indicator, but forced to result in a balanced design. 

This very simple setting is looked at from two different perspectives. From the traditional statistical inference point of view, we may be interested in testing for equality of means, i.e. testing $H_0: \beta_1^1 = 0$ vs. $H_1: \beta_1^1 \neq 0$ ($H_0$ agrees with $m^0$ and $H_1$ with $m^1$). However, we may also look at it from a prediction perspective: does the binary predictor $X$ improve the predictions of the outcome (i.e. is $m^1$ a better prediction model than $m^0$)?
In terms of significance testing, model $m^1$ is preferred over $m^0$ (and hence the binary $X$ is associated with the mean outcome) if the $p$-value is smaller than the nominal significance level. From the viewpoint of prediction modelling, $X$ can be considered an important predictor if model $m^1$ performs better than $m^0$ according to some prediction model assessment criterion (e.g. MSE). We will illustrate the use of the PIP as an inferential tool that can help in bridging the two worlds of statistical inference and prediction modelling. 


\subsubsection{Conditional PIP}
\label{S_2Sample_CondPIP}

Recall that the plug-in estimate of the conditional PIP coincides with the plug-in estimate of the theoretical PIP, where the unknown parameters in equation \eqref{pip_th} are replaced with their sample estimates. In the previous section, it was argued that simplifying the expression in \eqref{pip_th} depends on both the model parameters and $x^*$. Indeed, within the two-sample setting, Appendix \ref{A_theoretical} shows that the conditional PIP in the current two-sample setting is given by
\begin{align}
\label{cond_pip_general}
    p_\text{cond}(\hat{\mb\beta^0},\hat{\mb\beta^1},F^*) = \begin{cases} 
    0.5\ F^*_{Y^*|X^* =0}\left(\frac{\hat{\beta^0_0}+\hat{\beta^1_0}}{2}\right) + 0.5\left[1-F^*_{Y^*|X^* =1}\left(\frac{\hat{\beta^0_0}+\hat{\beta^1_0}+\hat{\beta^1_1}}{2}\right)\right] &  \text{ if } \hat{\beta^1_1} > 0, \\
  0.5\left[1-F^*_{Y^*|X^* =0}\left(\frac{\hat{\beta^0_0}+ \hat{\beta^1_0} }{2}\right)\right] + 0.5F^*_{Y^*|X^* =1}\left(\frac{\hat{\beta^0_0}+\hat{\beta^1_0}+\hat{\beta^1_1}}{2}\right)& \text{ if } \hat{\beta^1_1} < 0. \\
  \end{cases}
\end{align}

If we further assume that $F^*$ is a Gaussian distribution with $\E{Y^*\mid X^*=x}=m^1(\mb\beta^1,x)$ and $\var{Y^*\mid X^*=x}=\sigma_1^2$, then the plug-in estimate of the conditional PIP of Equation (\ref{cond_pip_general}) reduces to 
\begin{equation}
  p_{\text{cond};1}(\hat{\mb\beta^0},\hat{\mb\beta^1},\hat{F}^*,\hat{\mb\nu}) = \Phi\left(\frac{\lvert \hat{\beta_1^1}\rvert}{4\hat{\sigma}_1}\right),
\label{pip_C1}
\end{equation}
with $\Phi(\cdot)$ the distribution function of the standard normal distribution. This estimator will be referred to as C1. 

Upon using a Taylor series expansion (with $\sigma_1$ assumed to be known) and assuming that $\hat{\beta_1^1}$ is a consistent estimator of $\beta_1^1$, with $\beta_1^1 \neq 0$,
\begin{eqnarray*}
  p_{\text{cond};1}(\hat{\mb\beta^0},\hat{\mb\beta^1},\hat{F}^*,\hat{\mb\nu})
    &=& \Phi\left(\frac{\lvert {\beta_1^1}\rvert}{4\sigma_1}\right)
        + (\hat\beta_1^1 - \beta_1^1) \frac{1}{4\sigma_1} \phi\left(\frac{\lvert 
 {\beta_1^1}\rvert}{4\sigma_1}\right) \frac{\beta_1^1}{\lvert \beta_1^1\rvert} + o_P(n^{-1/2}).
\end{eqnarray*}
Hence, as $n\rightarrow \infty$,
\[
  p_{\text{cond};1}(\hat{\mb\beta^0},\hat{\mb\beta^1},\hat{F}^*,\hat{\mb\nu}) \convProb \Phi\left(\frac{\lvert\beta_1^1\rvert}{4\sigma_1}\right) = p_\text{th}(\mb\beta^0,\mb\beta^1,F^*,\mb{\nu}),
\]
and so the plug-in PIP estimator is consistent as well. As an additional result, we have shown that, as $n\rightarrow \infty$,
\[
   \sqrt{n}\left( p_{\text{cond};1}(\hat{\mb\beta^0},\hat{\mb\beta^1},\hat{F}^*,\hat{\mb\nu})-p_\text{th}(\mb\beta^0,\mb\beta^1,F^*,\mb{\nu}) \right) \convDistr Z \frac{1}{4\sigma_1} \phi\left(\frac{\lvert \beta_1^1 \rvert}{4\sigma_1}\right),
\]
where $Z$ is the limiting normal distribution of $\sqrt{n}(\hat{\beta_1^1} - \beta_1^1)$. See Appendix \ref{A_theoretical} for a proof. 

As an alternative to the strong distributional assumption made above, we can also consider the empirical cumulative distribution function $\hat{F}^{*,\text{emp}}_{Y|X=a}(y) = \frac{1}{n_a}\sum_{i=1}^{n_a}{\mb{1}_{Y_i < y |X=a}}$, where $a=0,1$ and $n_a=\frac{n}{2}$ in the current balanced design. This results into a second version of the plug-in conditional PIP, denoted by
\begin{align}
p_{\text{cond};2}(\hat{\mb\beta^0},\hat{\mb\beta^1},\hat{F}^*,\hat{\mb\nu}) = 
\begin{cases} 
\frac{1-\hat{F}^{*,\text{emp}}_{Y|X=1}\left(\frac{\hat{\beta^0_0}+\hat{\beta^1_0}+\hat{\beta^1_1}}{2}\right)}{2} + \frac{ \hat{F}^{*,\text{emp}}_{Y|X=0}\left(\frac{\hat{\beta^0_0}+\hat{\beta^1_0}}{2}\right)}{2} &  \text{ if } \hat{\beta^1_1} > 0, \\
\frac{1-\hat{F}^{*,\text{emp}}_{Y|X=0}\left(\frac{\hat{\beta^0_0}+\hat{\beta^1_0}}{2}\right)}{2} + \frac{ \hat{F}^{*,\text{emp}}_{Y|X =1}\left(\frac{\hat{\beta^0_0}+\hat{\beta^1_0}+\hat{\beta^1_1}}{2}\right)}{2}& \text{ if } \hat{\beta^1_1} < 0. \\
  \end{cases}
\end{align}
These estimators will be referred to as C1 and C2, respectively, and they are also plug-in estimators of the theoretical PIP. 

\subsubsection{Expected PIP}
\label{S_2Sample_ExpPIP}

The expected PIP requires the sampling distribution of the estimators of the regression coefficients. More specifically, we need to find the probability $\probf{X^*,Y^*}{E_1^2 < E_0^2}$ where $E_1 = m^1(\hat{\mb\beta}^1)-Y^*$ and $E_0 = m^0(\hat{\mb\beta}^0)-Y^*$ are two stochastically dependent variables. For the 2-sample setting we have a bivariate normal mixture distribution for $\mb{E}^t=(E_1,E_0)$:
\begin{equation}
\label{expected_simple_reg}
\mb{E} = \begin{pmatrix} 
E_1  \\
E_0
\end{pmatrix}\sim 0.5\mathcal{N}\left(
\begin{pmatrix} 
0  \\
-0.5\beta_1^1
\end{pmatrix},\sigma^2 \begin{pmatrix}
1+\frac{2}{n} & 1+\frac{1}{n} \\
1+\frac{1}{n} & 1+\frac{1}{n}
\end{pmatrix}\right)+ 0.5\mathcal{N}\left(
\begin{pmatrix} 
0  \\
0.5\beta_1^1
\end{pmatrix},\sigma^2 \begin{pmatrix}
1+\frac{2}{n} & 1+\frac{1}{n} \\
1+\frac{1}{n} & 1+\frac{1}{n}
\end{pmatrix}\right)    
\end{equation}
Monte Carlo sampling from this mixture distribution, with $\beta_1^1$ and $\sigma^2$ replaced with their estimates, is used to obtain a plug-in estimate of the expected PIP. A proof is given in Appendix \ref{A_expected}. 
For the more general linear regression case, where $X$ is continuous instead of binary, the results for both the conditional and expected PIP extend in a natural way as provided in Appendices \ref{A_theoretical} and \ref{A_expected}. 
Also for generalized linear models (e.g. logistic or Poisson regression), the conditional PIP could still easily be derived, starting from Equation (\ref{pip_th}) for the squared loss. However, the expected PIP may involve more complicated calculations for finding the bivariate distribution of $\mb{E}$. Nevertheless, the non-parametric approaches can always be applied, also with other loss functions.
    
\subsubsection{Relationship with p-values and MSE}
\label{S_2sample_PVal}

In the introduction the connection between the $p$-value (Equation (\ref{Eq_PValue})) and the conditional PIP was explained. In the simple case of the 2-sample problem, this link can be made explicit. First we write the $p$-value for the 2-sided test for testing $\beta_1^1=0$ as
\[
  p_p = \probf{T}{T \geq \frac{\lvert \hat\beta_1^1({\cal{O}}) \rvert}{\text{se}(\hat\beta_1)} \mid {\cal{O}}}
      = 2\left[1-F_{t;n-2}\left(\frac{\sqrt{n}\lvert \hat\beta_1^1({\cal{O}}) \rvert}{2\hat\sigma_1}\right)\right],
\]
where $T$ is $t$-distributed with $n-2$ degrees of freedom with distribution function  $F_{t;n-2}(\cdot)$. Upon using Equation (\ref{pip_C1}) as an expression for the C1 plug-in estimate of the conditional PIP, we find an explicit relationship,
\begin{equation}
\label{expr:estimator_reduced2}
 p_{\text{cond:1}}(\hat{\mb\beta^0},\hat{\mb\beta^1},\hat{F}^*,\hat{\mb\nu}) = \Phi\left(\frac{1}{2\sqrt{n}} F_{t;n-2}^{-1}(1-0.5 p_p)\right).
\end{equation}

\begin{figure}
    \centering
    \subfloat[]{{\includegraphics[width=4.3cm,height=3.2cm]{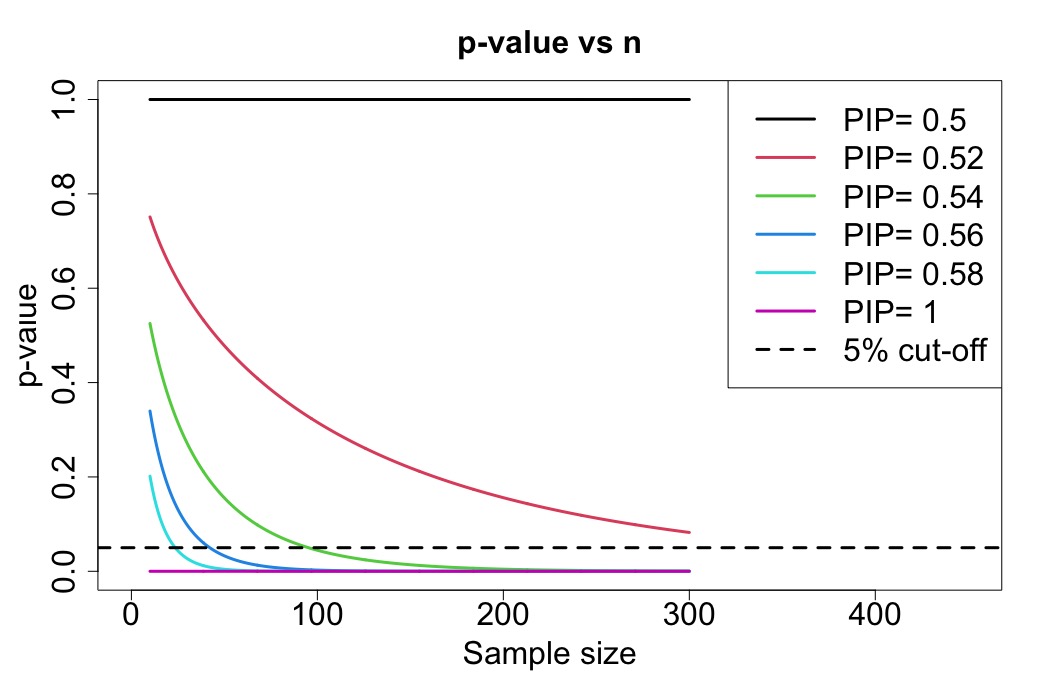} }}
    \qquad
    \subfloat[]{{\includegraphics[width=4.3cm,height=3.2cm]{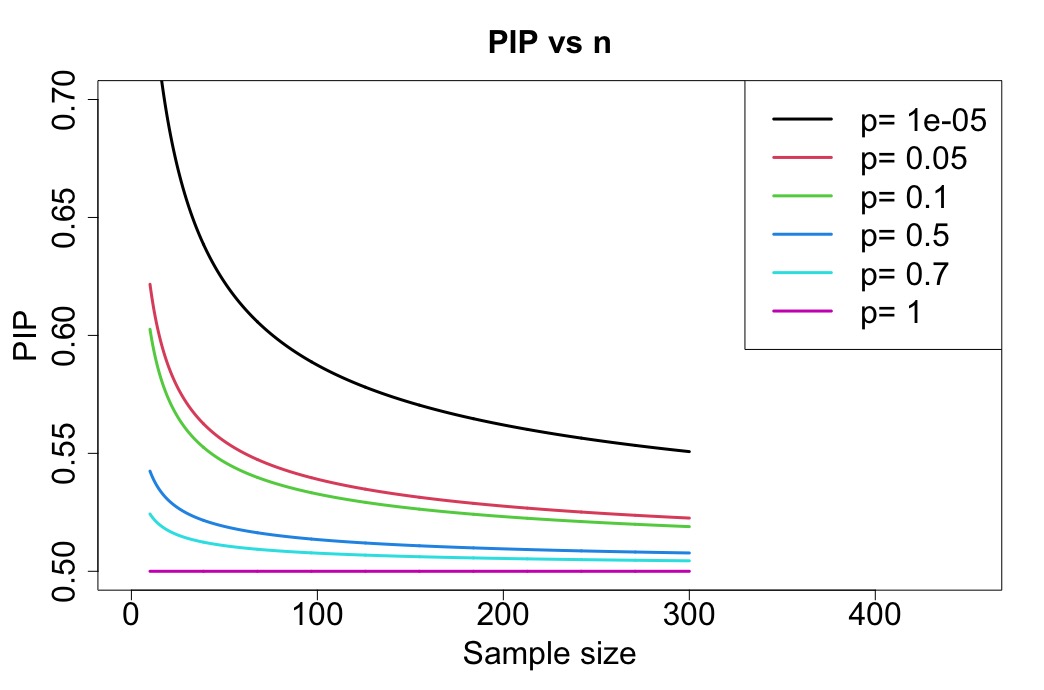} }}
    \qquad
    \subfloat[]{{\includegraphics[width=4.3cm,height=3.2cm]{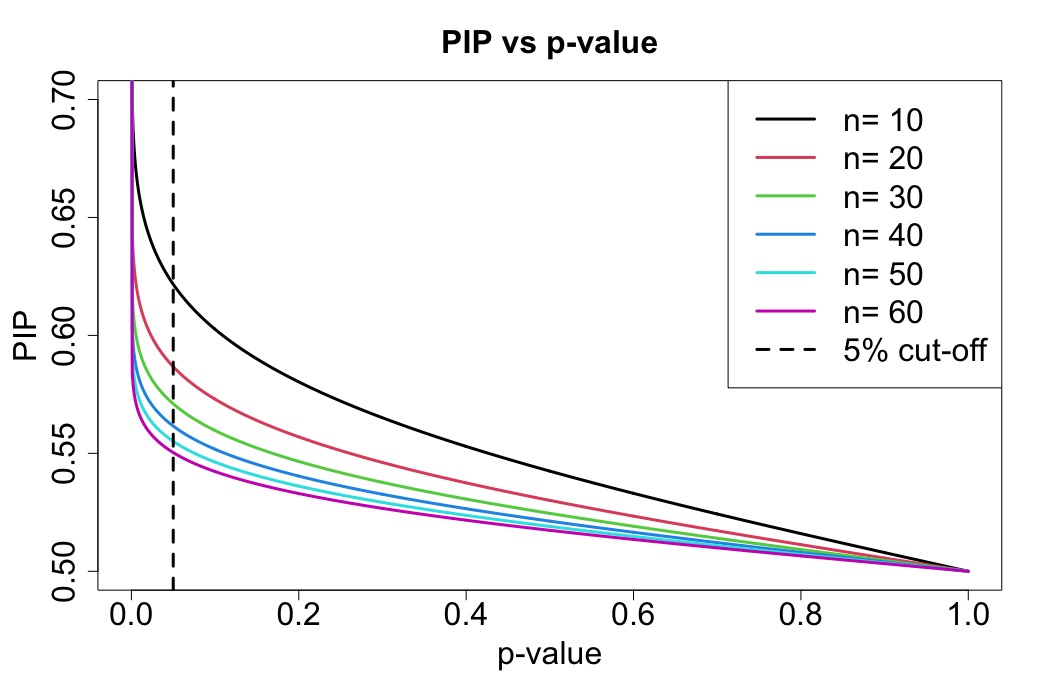} }}

    \caption{Relationships between the PIP as estimated by C1, the $p$-value and the sample size $n$ for the two-sample case. The reference lines in (a) and (c) correspond to $p_p=0.05$.} \label{Fig:PIP_p_n}
\end{figure}

Although both the $p$-value and the PIP estimate are statistics computed from the sample data, there are fundamental differences. The asymptotic behaviour of the $p$-value is different under the null and the alternative hypotheses. Under the null hypothesis the $p$-value asymptotically has uniform distribution (sometimes even for finite sample sizes), but under a fixed alternative the $p$-value of a consistent test converges in probability to zero. The conditional PIP, on the other hand, is a consistent estimator of a meaningful non-trivial probability. Despite these fundamental differences, for the two-sample problem we can establish an asymptotic relationship. In  Appendix \ref{A_relations} we show the following result for the $p$-value $p_n$ (i.e. the p-value based on sample size $n$): as $n\rightarrow \infty$,
\begin{equation}
  \label{eq:PAssymptotic}
  n^{-1} \ln p_n \convProb -\frac{1}{2}\ln\left(1+4\left(\Phi^{-1}\left(\text{PIP}\right)\right)^2\right),
\end{equation}
where the factor $n^{-1}$ on the left-hand side suggests that, although extremely small p-values are not informative for very large sample sizes, the PIP still has a relevant interpretation.

Figure \ref{Fig:PIP_p_n} shows plots for the relation between two of the three variables in Equation \ref{expr:estimator_reduced2}, while keeping the third fixed.  Panel (c) shows the relationship for fixed $n$, illustrating that for small $p$-values, still a large range of informative PIPs can be obtained, particularly for large $n$. Panel (a) shows that for a fixed PIP estimate, the $p$-values converge to zero with increasing sample sizes (except for the extreme PIP estimates of 1 and $0.5$). On the other hand, the relationship between the PIP estimate and the sample size $n$ in panel (b) demonstrates that the PIP estimates remain at non-trivial levels, away from $0.5$ (except when $p=1$). In the limit, however, as $n\rightarrow \infty$, the $p$-value cannot be considered fixed under fixed alternatives, and we need to resort to the asymptotic result of the scaled $p$-value (Equation \ref{eq:PAssymptotic}) to understand the relationship. 
For other, simple parametric models, similar explicit relationships between $p$-values and the PIP can be established, but for more complicated models and in the absence of distributional assumptions, often hypothesis tests and $p$-values are not available, where as the nonparametric PIP estimators do.

A second relation is observed with the difference in mean squared error, $\Delta$MSE, defined as the difference between the MSE of the full model and the MSE of the null model. More specifically, it is found that (see Appendix \ref{A_relations} for the derivation): 
\[
\widehat{\Delta MSE} = -4\sigma^2\Phi^{-1}(p_{\text{cond:1}}(\hat{\mb\beta^0},\hat{\mb\beta^1},\hat{F}^*,\hat{\mb\nu}))^2.
\]



Appendix \ref{A_relations} also shows a relation between the conditional PIP and the predictive distribution discussed by \cite{Clarke2012} and \cite{Billheimer2019}.


\section{Simulation study}
\label{sec:sim}
In this section, we first investigate the two-sample setting to better understand the behaviour of the three types of PIP and the underlying relation with the p-value and MSE. Next, a more complicated situation is presented to illustrate the general applicability of the nonparametric estimators.

\subsection{Two-sample setting}

In this simulation study, we have included the plug-in estimators C1 and C2 as described in Section \ref{S_2Sample_CondPIP}. These are estimators of both the theoretical and the conditional PIP. We have also included the plug-in estimator of the expected PIP (Exp) of Section \ref{S_2Sample_ExpPIP}. Among the non-parametric estimators (Section \ref{Nonpara}), we consider the leave-one-out (LOO), the 5-fold cross-validation (CV5) and the repeated 5-fold cross-validation (rep$\_$CV5) estimators. They will be evaluated as estimators of the expected and/or the conditional PIP. Finally, the split-sample (SS) estimator is included, again as an estimator of the theoretical and/or the conditional PIP.
Three effect sizes were considered ($\beta_1^1 = 0, -1$ and  $-4$) and the total sample sizes were 20, 40, 60, 100 and 400. For each scenario, 10000 simulation runs were performed. 

Results for the theoretical and expected PIP, and results for the conditional PIP are discussed separately. For the former two, the PIP remains constant throughout all simulations for they only depend on fixed parameters. The conditional PIP, on the other hand, depends on the sample and hence varies from simulation to simulation run.

\subsubsection{The Theoretical and Expected PIPs}

\begin{figure}
\centering
\begin{subfigure}[b]{0.85\textwidth}
   \includegraphics[width=1\linewidth]{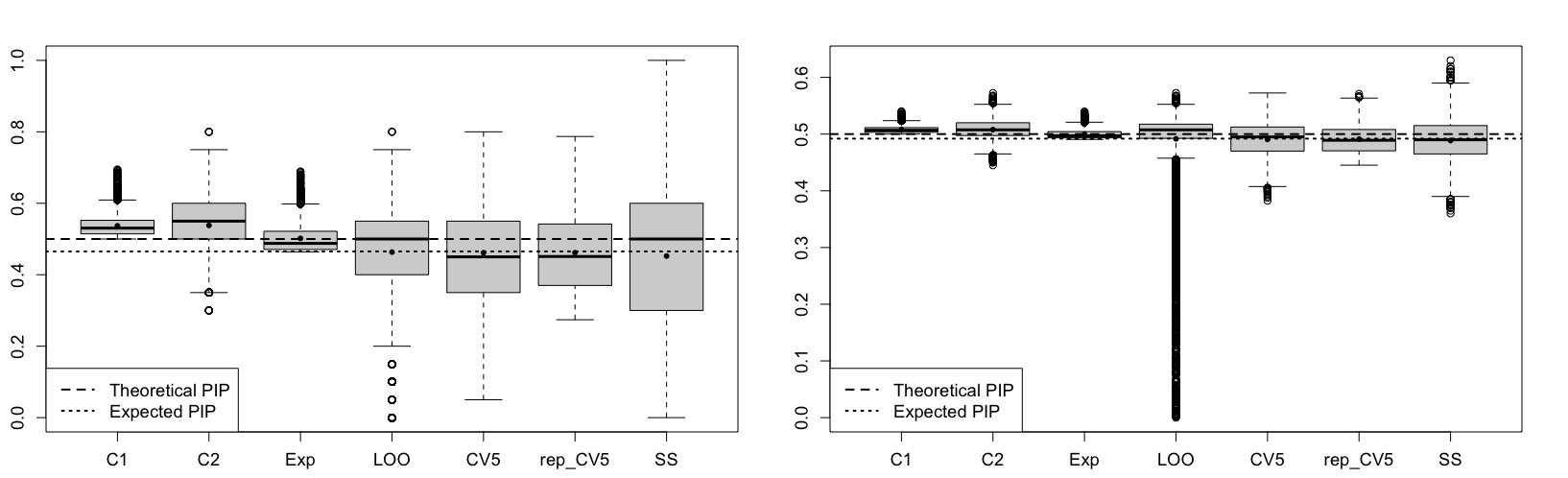}
   \caption{Effect size $\beta_1^1 = 0$ and sample sizes $n=20$ (left), $n=400$ (right)}
   \label{Sim0} 
\end{subfigure}

\begin{subfigure}[b]{0.85\textwidth}
   \includegraphics[width=1\linewidth]{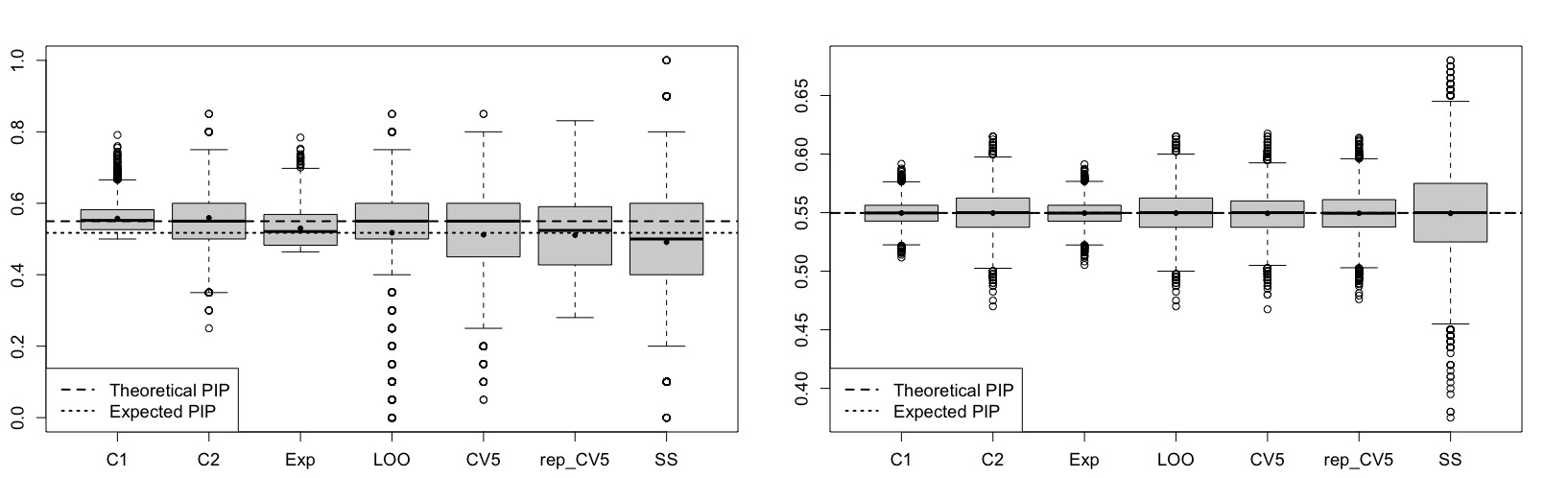}
   \caption{Effect size $\beta_1^1 = -1$ and sample sizes $n=20$ (left), $n=400$ (right)}
   \label{Sim1}
\end{subfigure}
 \begin{subfigure}[b]{0.85\textwidth}
   \includegraphics[width=1\linewidth]{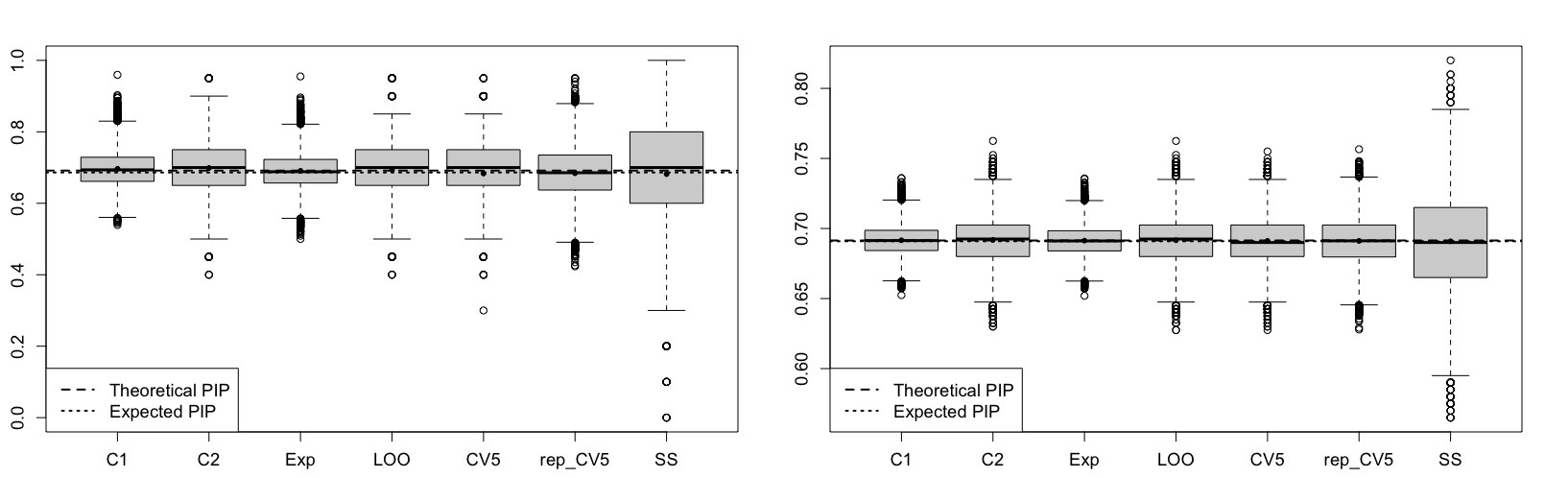}
   \caption{Effect size $\beta_1^1 = -4$ and sample sizes $n=20$ (left), $n=400$ (right)}
   \label{Sim4}
   
\end{subfigure}
  \caption{Results for 10000 simulation runs. $C1$ and $C2$ refer to $p_{\text{cond};1}(\hat{\mb\beta^0},\hat{\mb\beta^1},\hat{F}^*,\hat{\mb\nu})$ and $p_{\text{cond};2}(\hat{\mb\beta^0},\hat{\mb\beta^1},\hat{F}^*,\hat{\mb\nu})$, respectively; Exp refers to the plug-in estimate for the expected PIP; LOO, CV5, rep$\_$CV5 and SS refer to leave-one-out CV, 5-fold CV, repeated 5-fold cross-validation and split sampling, respectively. The true theoretical (dashed) and expected PIP (dotted) are indicated as horizontal reference lines. The mean over the simulation runs is indicated by the solid dot in the boxplot.} \label{fig:sim_th_exp}
\end{figure}

The results for the estimators of the theoretical and expected PIP are shown in Figure \ref{fig:sim_th_exp} for sample sizes 20 and 400. The plots for the other sample sizes are similar and can be found in Appendix \ref{C_Sims}. 

First we note that, as sample size increases, the true theoretical and expected PIPs converge to one another (see also Section \ref{S_RelationsBetweenPIPs}). The higher the effect size, the faster this convergence. Hence, for large sample sizes, the estimators discussed in this section, refer to both types of PIPs. 

In accordance with the overview of Figure \ref{fig_relationships}, we will consider all estimators as estimators of both the expected and theoretical PIP, except the resampling estimators (LOO, CV5 and repCV5) as estimators of the theoretical PIP.

First we focus on the results for $\beta_1^1\neq 0$, i.e. $m^1$ is a better model than $m^0$. From Figures \ref{Sim1} and \ref{Sim4} we conclude that all estimators are approximately unbiased, with a few exceptions. The plug-in estimators (C1, C2 and Exp) are slightly biased upwards for small sample sizes, and the split-sample (SS) estimator is slightly biased downwards, but the bias disappears when the sample size and/or the effect size $\beta_1^1$ increases. In terms of variance, the plug-in estimator C1 often shows the smallest variance; this is a highly parametric estimator and all model assumptions are satisfied in this set of simulations. The LOO and SS estimators show less desirable features: the LOO sampling distribution often shows long tails, and the SS estimator often has the largest variance. Our results suggest that the cross-validation estimators CV5 and rep$\_$CV5 are to be recommended as they have a good bias/variance trade-off. 

Turning to the results for the $\beta_1^1=0$ scenario ($m^0=m^1$). Overall the conclusions are very similar, but this scenario demonstrates an important shortcoming of the C1 plug-in estimator: in none of the simulation runs, the C1 estimate was smaller than $0.5$. In other words, the C1 estimates always suggest that $m^1$ is the better model, whereas the dummy regressor $X$ is not related to the outcome. Moreover, if $m^0$ is not worse than $m^1$, then $m^0$ should even be preferred over $m^1$, because (1) $X$ is not related to the outcome (as in $m^0$), and (2) $m^0$ does not suffer from overfitting. The explanation for this phenomenon is that in C1 the distribution is set to $F^*$, which is in turn set to $m^1$. In both $F^*$ and $m^1$ the same estimate of $\beta_1^1$ is used and hence, according to C1, $m^1$ will never perform worse than $m^0$.

\subsubsection{The Conditional PIP}

Because for a given sample, the sample estimate $\hat{\mb\beta}^1$, which is part of $m^1$, does not coincide with the true $\mb\beta^1$, which is part of the distribution $F^*$, the latter no longer agrees with $m^1$, and the analytical computation of the true conditional PIP is less straightforward than before. We therefore approximate this PIP by an average of $n_t=1\times 10^6$ randomly generated realisations $(x_i^*,y_i^*)$ from $F^*$. In particular, 
\begin{equation*}
    \frac{1}{n_t} \sum_{i=1}^{n_t} \indicat{L(m^1(\hat{\mb\beta}^1,x_i^*),y_i^*) < L(m^0(\hat{\mb\beta}^0,x_i^*),y_i^*)}.
\end{equation*}

For the conditional PIP, only the plug-in estimator of the expected PIP is not considered as a potential estimator (see Figure \ref{fig_relationships}). From a theoretical perspective \citep{Bates2021}, the resampling based estimators target the expected PIP, but we still also evaluate them for estimating the conditional PIP; recall that for large sample sizes both PIPs are very similar.

\begin{figure}
\centering
\begin{subfigure}[b]{0.85\textwidth}
   \includegraphics[width=1\linewidth]{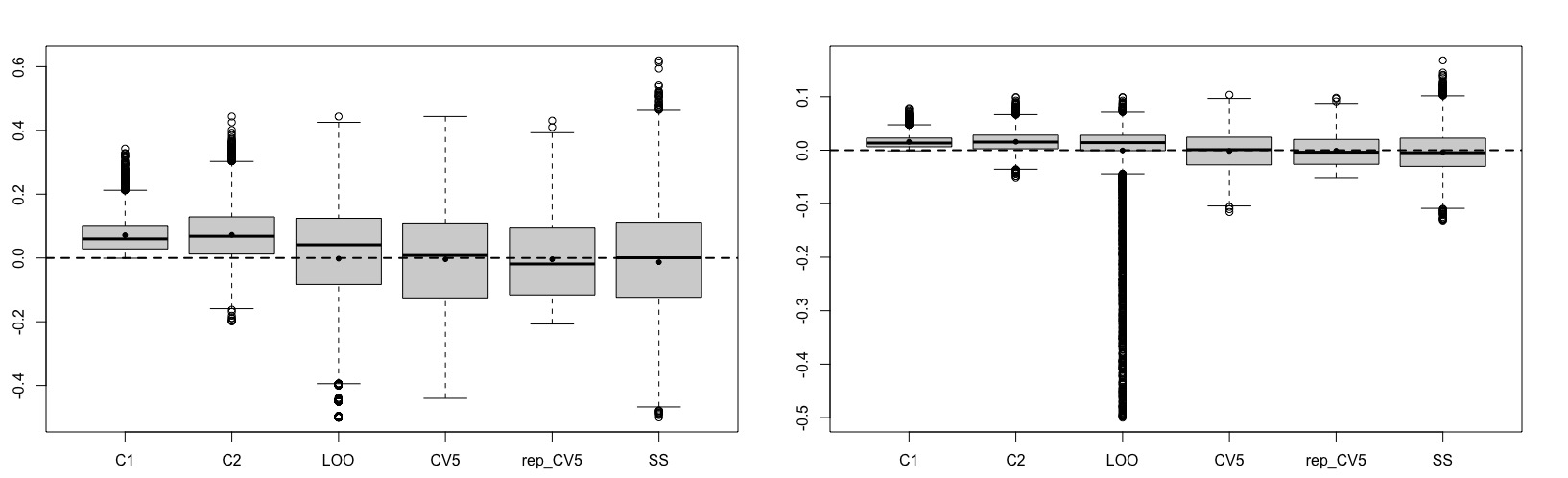}
   \caption{Effect size $\beta_1^1 = 0$ and sample sizes $n=20$ (left), $n=400$ (right)}
   \label{Sim0_comp} 
\end{subfigure}

\begin{subfigure}[b]{0.85\textwidth}
   \includegraphics[width=1\linewidth]{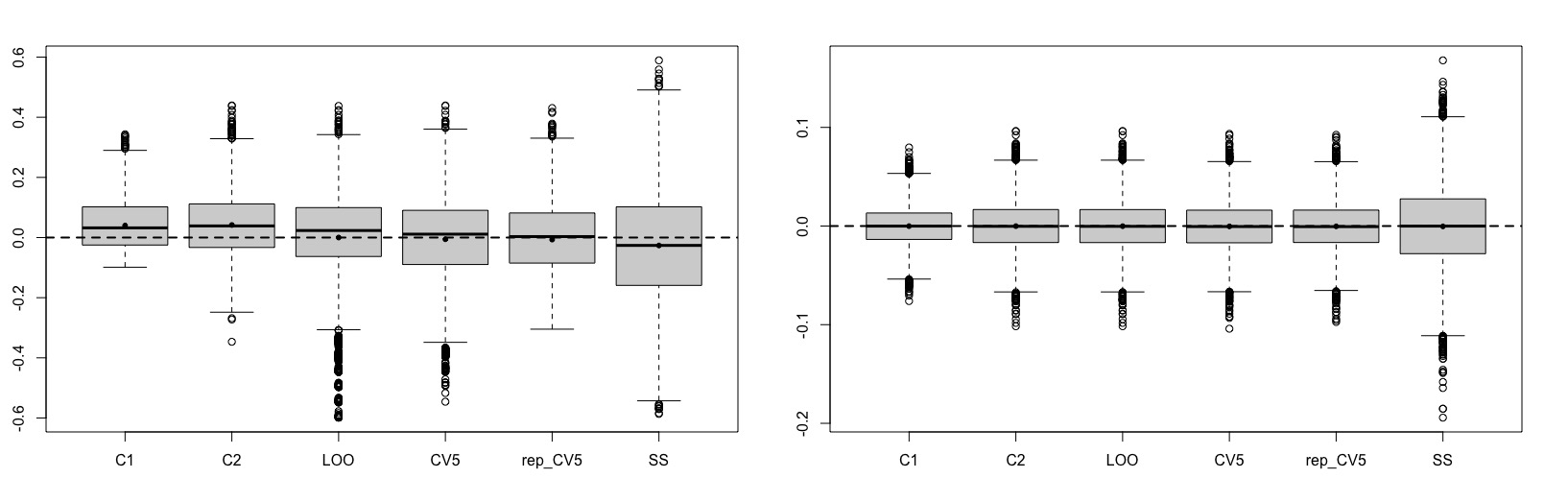}
   \caption{Effect size $\beta_1^1 = -1$ and sample sizes $n=20$ (left), $n=400$ (right)}
   \label{Sim1_comp}
\end{subfigure}
 \begin{subfigure}[b]{0.85\textwidth}
   \includegraphics[width=1\linewidth]{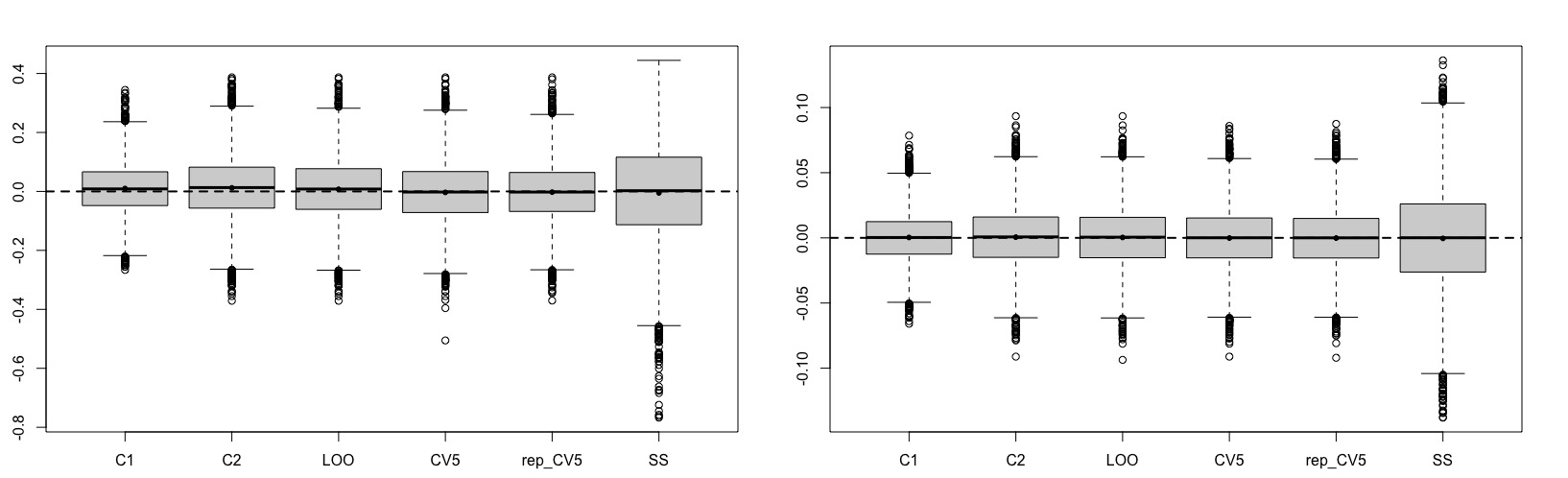}
   \caption{Effect size $\beta_1^1 = -4$ and sample sizes $n=20$ (left), $n=400$ (right)}
   \label{Sim4_comp}
   
\end{subfigure}
  \caption{Comparing the difference between the empirical conditional PIP and the PIP estimators. Similar abbreviations are used as compared to Figure \ref{fig:sim_th_exp}} \label{fig:sim_cond}.
\end{figure}

Figure \ref{fig:sim_cond} suggests that all nonparametric estimators are approximately unbiased. The plug-in estimators C1 and C2 show a small upward bias which becomes small as the sample size and/or the effect size increases. Just like before, the C1 estimates are never smaller than $0.5$ in the $\beta_1^1=0$ scenario. 
In terms of variance, the LOO and SS estimators perform worst. The smallest variance is obtained with C1 and C2, but since they suffer from a bias, we prefer the CV5 and rep$\_$CV5 estimators, which show a good bias/variance trade-off. Moreover, these nonparametric methods can be easily used in more complicated setting with e.g. nonlinear models. 



\subsubsection{The Relationships between PIP, P-values and MSE}

\begin{figure}
\centering
\begin{subfigure}[b]{0.85\textwidth}
   \includegraphics[width=1\linewidth]{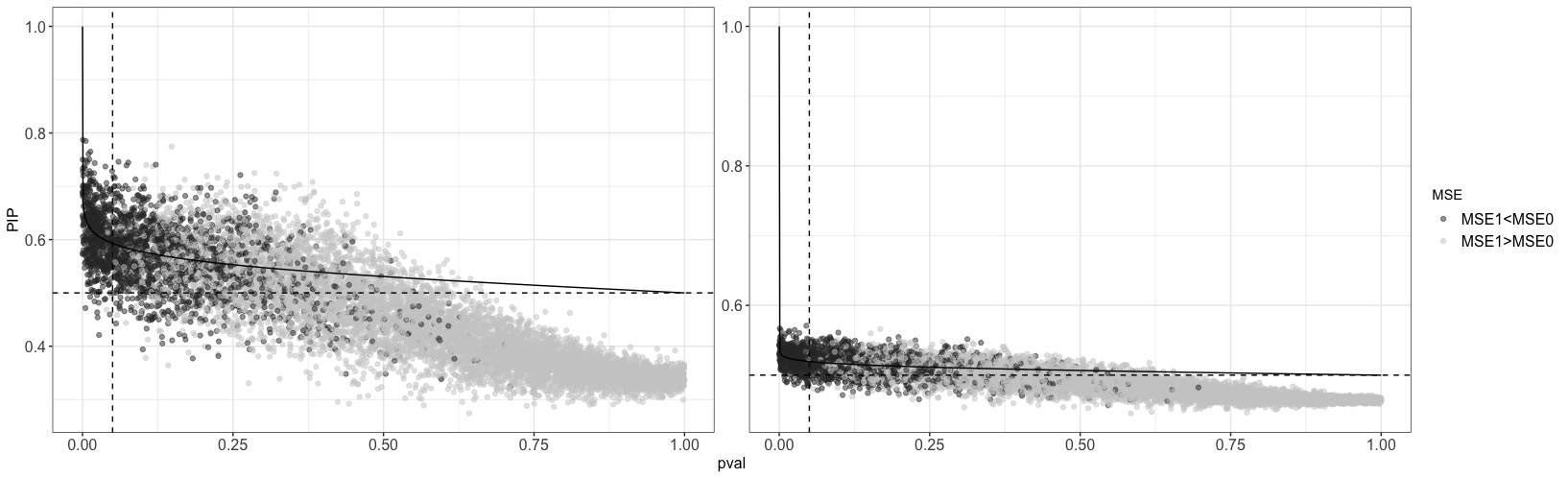}
   \caption{Effect size $\beta_1^1 = 0$ and sample sizes $n=20$ (left), $n=400$ (right)}
   \label{Sim0_rel} 
\end{subfigure}

\begin{subfigure}[b]{0.85\textwidth}
   \includegraphics[width=1\linewidth]{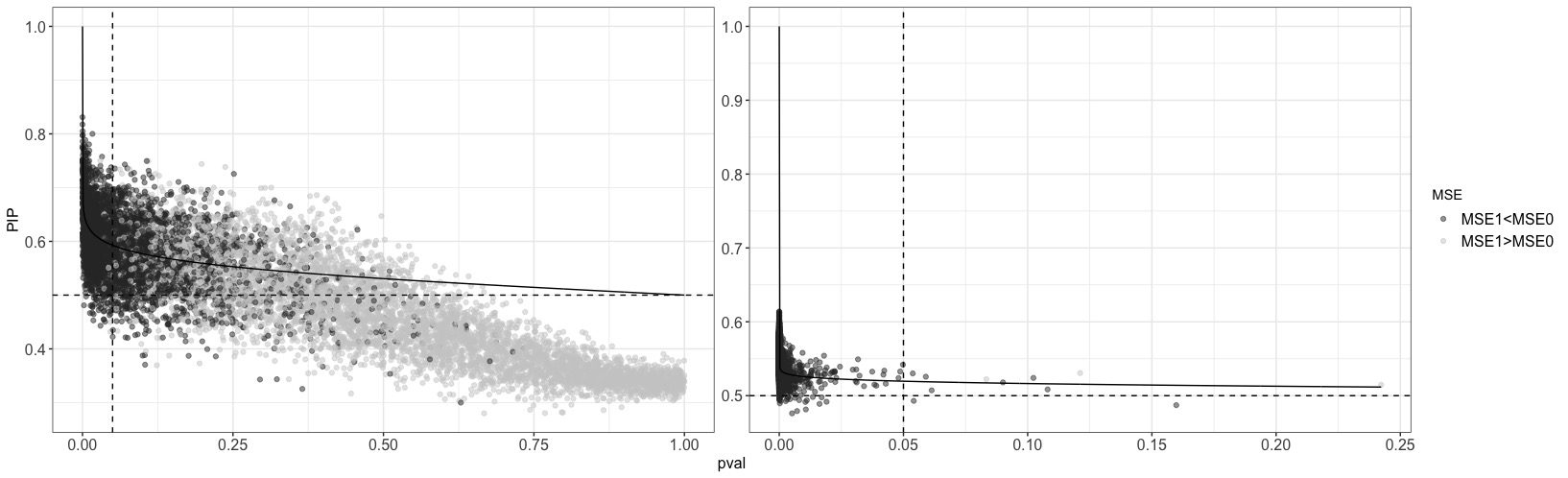}
   \caption{Effect size $\beta_1^1 = -1$ and sample sizes $n=20$ (left), $n=400$ (right)}
   \label{Sim1_rel}
\end{subfigure}
 \begin{subfigure}[b]{0.85\textwidth}
   \includegraphics[width=1\linewidth]{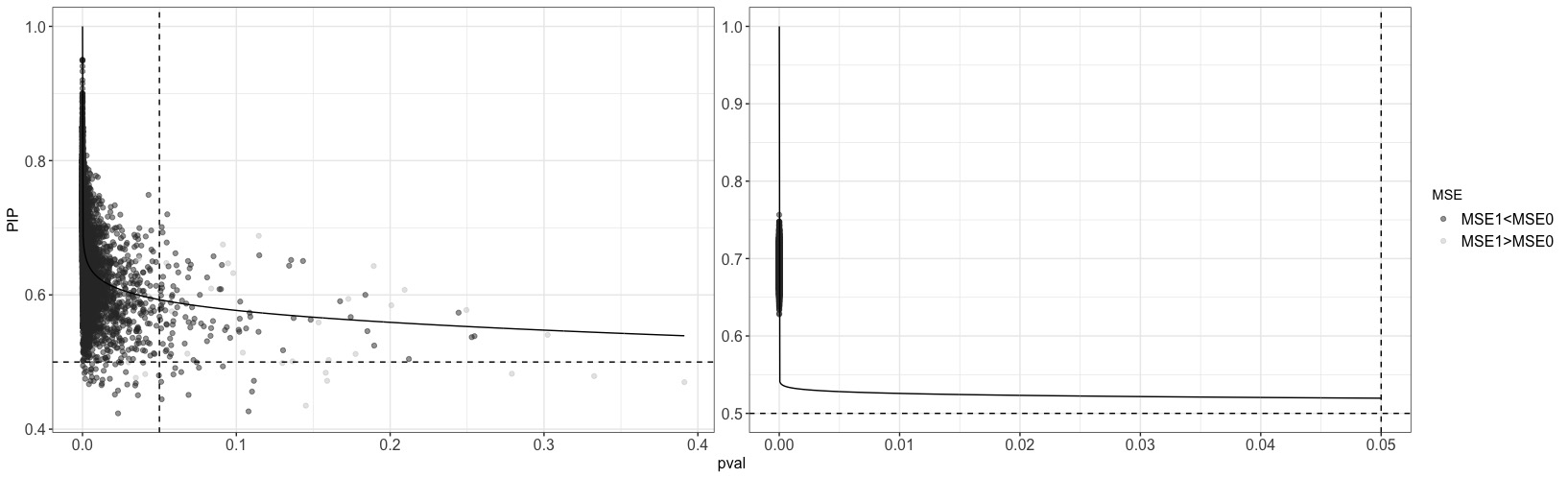}
   \caption{Effect size $\beta_1^1 = -4$ and sample sizes $n=20$ (left), $n=400$ (right)}
   \label{Sim4_rel}
   
\end{subfigure}
  \caption{Relation between repeated 5-fold CV PIP and the p-value for testing $\beta_1^1 = 0$. The gray scale reflects the difference between the MSE of $m^1$ and $m^0$. The solid line indicates the exact relation of \eqref{expr:estimator_reduced2}. The dashed reference lines refer to PIP = 0.5 and p-value = $0.05$.} \label{fig:rel}
\end{figure}

In Section \ref{S_2sample_PVal} the exact relationship between the p-value and the C1 plug-in estimator of the conditional PIP was investigated. For the nonparametric PIP estimators, on the other hand, such exact relationships do not exist. In this section we empirically investigate this relationship for the 5-fold CV estimator, in a simulation study. The results in Figure \ref{fig:rel} still show many features of the earlier exact relationship. The random scattering is a consequence of the random splitting of the dataset into the folds. 

The graphs show that the higher the p-value, the lower the PIP. Insignificant test results often correspond to a PIP smaller than $0.5$, which is an indication that $m^1$ overfits the data. For this simple 2-sample problem, the conclusions based on MSE and PIP often agree. Indeed, Table \ref{Tab:sim_pval_pip_mse} shows the percentages of "correct decisions" based on the p-value, the MSE difference and the repeated 5-fold CV PIP. For the latter, both the estimate (i.e. PIP) as well as the lower $95\%$ confidence bound (i.e. PIP$_{LB}$) are considered. The results confirm the fact that for the medium effect size and small sample sizes, both PIP and PIP$_{LB}$ outperform the p-value in terms of power and show comparable performance as the MSE decisions, with some underperformance for PIP$_{LB}$ and overperformance for PIP.  On the other hand, MSE, PIP and PIP$_{LB}$ seem to be too liberal when the effect size is 0, as compared to the p-value. All approaches show similar performance for the largest effect size. We want to stress, however, that it is not our intention to advocate the use of sharp thresholds for decision making. We have used it here for making the comparisons easier.

\begin{table}
\centering
\caption{Comparison between decisions based on the p-value, the MSE difference between models $m^1$ and $m^0$ (denoted by $\Delta$ MSE), the repeated cross-validation PIP and its corresponding lower bound (denoted by PIP and PIP$_{LB}$, respectively). The table shows the percentages of a correct decision}.
\begin{tabular}{@{\extracolsep{4pt}}ccccccccc @{}}
 \hline
 &\multicolumn{4}{c}{$\beta_1^1=0$}&\multicolumn{4}{c}{$\beta_1^1=-1$} \\
 \cline{2-5} \cline{6-9}  
  Sample size & p-value&$\Delta$ MSE&PIP$_{LB}$&PIP& p-value&$\Delta$ MSE&PIP$_{LB}$&PIP\\
 \hline
20& 94.81 & 80.05 & 84.40&63.48 &18.30 & 41.13 & 31.62&57.59 \\
40&95.02 & 79.44 & 84.92&63.47 &33.88 & 56.57 & 46.18&70.45  \\
60&94.92 & 79.15 & 84.15&64.22 &48.22 & 68.32 & 58.83& 79.30 \\
100&94.96 & 79.16 & 83.80&64.48 &70.67 & 84.15 & 76.72&89.78  \\
400&94.98 & 79.71 & 82.11&64.46 &99.89 & 99.89 & 99.20&99.72 \\
\hline
\end{tabular}
\label{Tab:sim_pval_pip_mse}
\end{table}

\subsection{Nonlinear models and gradient boosting machines}
\label{Nonlinear}

The nonparametric estimators are also applicable in more complicated settings. The results in this section originate from  data that were sampled from a nonlinear model of the form $y=\left|4x_1\right|^{3x_4}+5x_2+(2x_3)^{x_5} +\varepsilon$, with given predictors ($x_1$ up to $x_5$) and zero-mean error term $\varepsilon$.
More details on the data generating mechanism and the gradient boosting machines are provided in Appendix \ref{C_Sims}. In short, the different estimators of the PIP are evaluated for comparing two models estimated with gradient boosting machines: the full model $m^1$ makes use of all covariates, while the null model $m^0$ only considers variables $x_1$, $x_2$ and $x_3$. This comes down to testing whether nonlinear effects are required. Estimates of the PIP are obtained based on 5-fold cross-validation, 10 times repeated 5-fold cross-validation and split sampling. In addition to the PIP, the difference in MSE between both models was calculated as well, where the MSE is estimated based on both types of cross-validation. In the absence of formal p-values for gradient boosting machines, the aim is to use the PIP to provide information about the importance of covariates $x_4$ and $x_5$. Table \ref{Tab:gbm} summarises the results in terms the percentage of correct decisions made by PIP and MSE when choosing between models $m^1$ and $m^0$ (i.e. percentage of runs where PIP$>$0.5 and MSE1$<$MSE0).

\begin{table}
\centering
\caption{Comparison between decisions based on the different nonparametric PIP estimators and the MSE difference between models $m^1$ and $m^0$ (denoted by $\Delta$ MSE). The table shows the percentages of a correct decision.}
\begin{tabular}{@{\extracolsep{4pt}}cccccc@{}}
 \hline
 &\multicolumn{3}{c}{PIP}&\multicolumn{2}{c}{$\Delta$ MSE} \\
 \cline{2-4} \cline{5-6}  
  Sample size &SS& CV5 & rep$\_$CV5& CV5&rep$\_$CV5\\
 \hline
40&75.00 & 92.36 & 94.05&  99.07 &98.99 \\
60&84.31&98.19  & 98.99  & 99.76   &   99.70  \\
100&95.06&99.95  & 99.95  &99.97 & 99.94\\
400&100.00  & 100.00 &100.00 &100.00 &100.00\\
\hline
\end{tabular}
\label{Tab:gbm}
\end{table}
 
The higher the sample size, the better the agreement between decisions based on MSE and those based on PIP, but already for the small sample sizes the performances are quite similar. As expected, the (repeated) CV PIP shows better performance as compared to split sampling, with already great performance for the smallest sample size under consideration. Of course, it is noted that the results on this more complicated setting are quite limited, but they already indicate the ease of applicability of the PIP and hint in the direction of a similar performance when compared to e.g. MSE. In addition, the absence of p-values for these type of machine learning models can be (partly) resolved by looking at the PIP.    

\section{Application}
\label{sec:app}

\cite{Camerer2018} evaluated the replicability of 21 social science experiments published in the journals Nature and Science between 2010 and 2015, each of which reported a significant p-value in the original report. In addition to obtaining insights from experts in the field through the use of market predictions and surveys, the team of authors also collected new data to check whether these supported the original conclusions. Needless to say, this was a daunting and time consuming task. In absence of the true original data sets, we used the reported parameter settings of four studies for simulating our own data. Our settings are listed in Appendix \ref{D_Application} and they were chosen such that they approximate the original p-values as close as possible.   
Table \ref{tab:application} shows results from these four studies as presented in \cite{Camerer2018}, who found two studies to be replicable and two not. The table also shows the estimated PIP (rep\_CV5), with a lower and upper confidence bound. The PIP is an estimate of whether the model including the original significant effect provides a better prediction than the model without that effect. In the first study, the PIP is close to 0.5 and the lower bound is even below 0.5, pointing towards an overfitting model and hence low belief that the effect is actually significant. Similarly, the PIP and corresponding lower bound in the second study are also close to 0.5. In combination with a boundary p-value, replicability seems questionable. In contrast, the PIP and lower confidence limit are higher than 0.63 in studies three and four, which increases the belief that the effect could be replicated. 

In conclusion, with only a minor additional computational effort, the PIP can be used in combination with the p-value to provide more details on the strength of the conclusions that would normally be based on the p-value alone. Herewith, the PIP fits nicely within the suggestions made by \cite{Wasserstein2019} to overcome some of the concerns related to p-values as discussed by e.g. \cite{Amrhein2019}. Notwithstanding our focus here on the PIP and p-values, we evidently recommend to also report effect size estimates and their confidence intervals as an essential part of the data analysis. 

\begin{table}[htbp]
\caption{Application of the PIP (rep\_CV5) for four studies discussed in \cite{Camerer2018}. Columns 4 (PredMarket) and 5 (Survey) report the results from the peers and should be interpreted as the probability that experts assign to a possible replication of the results \label{tab:application}}
\begin{center}
\footnotesize
\begin{tabular}{rccccccc}
Study & p-value & Replicated & PredMarket & Survey &PIP$\_$lower&PIP&PIP$\_$upper\\\hline
\cite{Gervais2012} & 0.0291 & No & 0.1700 & 0.2000&0.4913&0.5207&0.5470 \\
\cite{Ackerman2010} & 0.0488 & No & 0.1500 & 0.1300 &0.5183&0.5624&0.5994 \\
\cite{Balafoutas2012} & 0.0194 & Yes & 0.7500 & 0.4300&0.6339&0.6389&0.6434 \\
\cite{Wilson2014} & $<$0.0001 & Yes & 0.4600 & 0.5200&0.6667&0.7140&0.7333 \\
\end{tabular}
\end{center}
\end{table}

\section{Conclusion}
\label{sec:conc}

We have introduced the probability of improved prediction (PIP) as an addition to the toolbox for statistical inference. In particular, we position the new concept somewhere in between hypothesis testing ($p$-values) and prediction modelling (e.g. the mean squared error (MSE) as a model selection criterion). The PIP is the probability that one model provides a better prediction as compared to another model, where "better" is based on a user-defined loss function. Three versions have been proposed: (1) the {\em theoretical PIP} does not depend on the data and assumes that all model parameters are known; (2) the {\em conditional PIP} makes use of parameter estimates, and is to be interpreted conditional on the observed sample data; and (3) the {\em expected PIP}, which is defined as the average of the conditional PIP over repeated samples. It is the conditional PIP for which we have established an explicit relationship with on the one hand the $p$-value, and, on the other hand, model selection based on the MSE. To gain further insight into these relationships, we have worked out details for the very simple setting of the two-sample problem (and linear models). Several estimators of the various PIPs have been proposed and evaluated in a simulation study. Plug-in estimators are straightforward for simple parametric models that come with explicit distributional assumptions, but they cannot be generalised to settings in which complicated nonlinear models are used (e.g. machine learning) or when no distributional assumptions can be made. For the sake of the latter situations, we have also proposed nonparametric PIP estimators based on sample splitting and cross validation. Simulation results suggest that the repeated 5-fold cross-validation estimator performs best as an estimator of the expected PIP. Note that cross-validation estimators are known to rather estimate the expected values (averaged over repeated samples) of targeted estimand, then that they estimate the estimand itself; see e.g. \cite{Bates2021} for a detailed discussion in the context of cross-validation in prediction modelling. Nevertheless, the simulation results also show that repeated cross-validation estimator only shows a small bias for estimating the conditional PIP, which is often considered to be the most relevant version.  

Our aim of introducing the PIP was to form a bridge between statistical inference and prediction modeling. Just like $p$-values, it has a probabilistic interpretation, but it has some advantages. First, for non-statisticians it has an easier interpretation than $p$-values, and it is therefore less prone to misinterpretation and misuse. Second, the PIP behaves as an estimator and converges to a meaningful probability, whereas the $p$-value, under a fixed alternative hypothesis, always converges to zero. As a consequence, the $p$-value hardly tells anything about the scientific relevance. Finally, in contrast to $p$-values, the PIP can be easily computed for many more models, even when computer intensive machine learning methods are used for model building. 
The PIP is also closely related to the MSE for comparing prediction models. Whereas the latter measures to what extent one model is better than the other {\em on average}, in terms of the squared prediction error, the PIP tells us how often the one model gives a better prediction than the other model. 
In this sense, we believe that the PIP can be an interesting addition to the toolbox of a statistician or data scientist.

\bibliographystyle{chicago}

\bibliography{References}

\newpage

\bigskip
\begin{center}
{\large\bf SUPPLEMENTARY MATERIAL}
\end{center}
\appendix

\section{Derivations related to the theoretical PIP}
\label{A_theoretical}

For ease of notation, the dependence of models $m^1$ and $m^0$ on $\mb{X}$ will be omitted and hence $m^i(\mb\beta^i) \equiv m^i(\mb\beta^i,\mb{X^*})$ for $i=0,1$. In case $m^1(\mb\beta^1) = m^0(\mb\beta^0)$, the theoretical PIP is defined to be 0.5. For the case where $m^1(\mb\beta^1) \neq m^0(\mb\beta^0)$ and using the squared error loss, it is derived that   
\begin{align}
p_\text{th}(\mb\beta^0,\mb\beta^1,F^*,\mb{\nu})&=\probf{X^*,Y^*}{L(m^1(\mb\beta^1),Y^*) < L(m^0(\mb\beta^0),Y^*)}  \nonumber \\
&=\probf{X^*,Y^*}{(m^1(\mb\beta^1)-Y^*)^2 < (m^0(\mb\beta^0)-Y^*)^2} \nonumber \\
&=\probf{X^*,Y^*}{m^1(\mb\beta^1)^2-2m^1(\mb\beta^1)Y^*+{Y^*}^2 <  m^0(\mb\beta^1)^2-2m^0(\mb\beta^0)Y^*+{Y^*}^2} \nonumber\\ 
&= \probf{X^*,Y^*}{m^1(\mb\beta^1)^2- m^0(\mb\beta^0)^2 < 2Y^*(m^1(\mb\beta^1)-m^0(\mb\beta^0)} \nonumber\\
&=
\int_{\cal{X^*}} \probf{Y^*|X^*}{m^1(\mb\beta^1)^2- m^0(\mb\beta^0)^2 < 2Y^*(m^1(\mb\beta^1)-m^0(\mb\beta^0))|X^*=x^*}dF^*_{X^*}(x^*),
\end{align}
with ${\cal{X^*}}$ the sampling space of $X^*$.

The further expansion of equation \eqref{pip_th} depends on dividing the sampling space into regions where the sign of $m^1(\mb\beta^1)-m^0(\mb\beta^0)$ is either positive or negative. However, this requires information on both the model parameters (i.e. $\mb\beta^0$ and $\mb\beta^1$) as well as on the value of $x^*$. A general expression  is therefore hard to derive, but for the case of simple linear regression with $m^0 = \beta_0^0$ and $m^1 = \beta_0^1 + \beta_1^1x$, simplified expressions can be obtained.

First of all, a link between the regression parameters can be easily obtained. Indeed, we know that 
\begin{equation*}
    E[Y] = \int_{\cal{X}} E[Y|X=x] dF_{X}(x).
\end{equation*}
Considering either of the two models $m^0$ and $m^1$ for $E[Y|X]$ should leave the unconditional expectation $E[Y]$ unchanged and hence
\begin{eqnarray*}
    E[Y] &=& \int_{\cal{X}} m^0(\mb{\beta^0},x) dF_{X}(x) \\
    &=& \int_{\cal{X}} m^1(\mb{\beta^1},x) dF_{X}(x)
\end{eqnarray*}
leads to
\begin{eqnarray*}
   \int_{\cal{X}} m^0(\mb{\beta^0},x) dF_{X}(x) -
 \int_{\cal{X}} m^1(\mb{\beta^1},x) dF_{X}(x) &=& 0 \iff \\
 \int_{\cal{X}} \left[m^0(\mb{\beta^0},x) -  m^1(\mb{\beta^1},x)\right] dF_{X}(x) &=& 0 \iff \\
  \int_{\cal{X}} \left[m^0(\mb{\beta^0},x) -  m^1(\mb{\beta^1},x)\right] f_{X}(x) dx &=& 0, 
\end{eqnarray*}
which in the case of simple linear regression simplifies to
\begin{eqnarray*}
    \int_{\cal{X}} \left[\beta_0^0 -  \beta_0^1 - \beta_1^1 x\right] f_{X^*}(x) dx &=& 0 \iff \\
 \beta_0^0  =  \beta_0^1 + E[X]\beta_1^1.
\end{eqnarray*}
Note that these final two steps become harder and sometimes impossible to solve when models are non-linear in the parameters, in x or both.

Now that the link between $\mb{\beta^0}$ and $\mb{\beta^1}$ has been established, it is further seen that
\begin{align*}
m^1(\mb\beta^1)-m^0(\mb\beta^0) &= \beta_0^1 + \beta_1^1 x^*  - \beta_0^1 - E[X^*]\beta_1^1 \\
&= \beta_1^1 (x^* - E[X^*] ).
\end{align*}
The sign of this expression is clearly related to the sign of $\beta_1^1$ and to the location of $x$ with respect to the expected value $E[X]$. More specifically, depending on the sign of $\beta_1^1$, the following situations can be distinguished:
\begin{itemize}
    \item if $\beta_1^1<0$:
    \[
    m^1(\mb\beta^1)-m^0(\mb\beta^0) > 0 \iff x^* < E[X^*]
    \]
    and putting ${\cal{X^*}} = {\cal{X}}^*_1 \cup {\cal{X}}^*_2 = \{x^* < E[X^*]\} \cup \{x^* > E[X^*]\}$, equation \eqref{pip_th} becomes
    \begin{align*}
     &&\int_{x^* < E[X^*]} \probf{Y^*|X^*}{\frac{m^1(\mb\beta^1)^2- m^0(\mb\beta^0)^2}{2(m^1(\mb\beta^1)-m^0(\mb\beta^0))} < Y^*|X^*=x^*}dF^*_{X^*}(x^*) + \\ 
     &&\int_{x^* > E[X^*]} \probf{Y^*|X^*}{\frac{m^1(\mb\beta^1)^2- m^0(\mb\beta^0)^2}{2(m^1(\mb\beta^1)-m^0(\mb\beta^0))} > Y^*|X^*=x^*}dF^*_{X^*}(x^*)   
    \end{align*}
   which further simplifies to
       \begin{align*}
     &&\int_{x^* < E[X^*]} \probf{Y^*|X^*}{\frac{m^1(\mb\beta^1)+ m^0(\mb\beta^0)}{2} < Y^*|X^*=x^*}dF^*_{X^*}(x^*) + \\ 
     &&\int_{x^* > E[X^*]} \probf{Y^*|X^*}{\frac{m^1(\mb\beta^1)+ m^0(\mb\beta^0)}{2} > Y^*|X^*=x^*}dF^*_{X^*}(x^*)   
    \end{align*}
    
    \item if $\beta_1^1>0$:
    \[
     m^1(\mb\beta^1)-m^0(\mb\beta^0) > 0 \iff x^* > E[X^*]
    \]
    and putting ${\cal{X^*}} = {\cal{X}}^*_1 \cup {\cal{X}}^*_2 = \{x^* < E[X^*]\} \cup \{x^* > E[X^*]\}$, equation \eqref{pip_th} becomes
    \begin{align*}
     &&\int_{x^* < E[X^*]} \probf{Y^*|X^*}{\frac{m^1(\mb\beta^1)^2- m^0(\mb\beta^0)^2}{2(m^1(\mb\beta^1)-m^0(\mb\beta^0))} > Y^*|X^*=x^*}dF^*_{X^*}(x^*) + \\ 
     &&\int_{x^* > E[X^*]} \probf{Y^*|X^*}{\frac{m^1(\mb\beta^1)^2- m^0(\mb\beta^0)^2}{2(m^1(\mb\beta^1)-m^0(\mb\beta^0))} < Y^*|X^*=x^*}dF^*_{X^*}(x^*)    
    \end{align*}
    which further simplifies to
       \begin{align*}
     &&\int_{x^* < E[X^*]} \probf{Y^*|X^*}{\frac{m^1(\mb\beta^1)+ m^0(\mb\beta^0)}{2} > Y^*|X^*=x^*}dF^*_{X^*}(x^*) + \\ 
     &&\int_{x^* > E[X^*]} \probf{Y^*|X^*}{\frac{m^1(\mb\beta^1)+ m^0(\mb\beta^0)}{2} < Y^*|X^*=x^*}dF^*_{X^*}(x^*)   
    \end{align*}
\end{itemize}  

\subsection{Binary covariate}

For the two-sample case study, a balanced design is considered for the binary $X^*$, meaning that $E[X^*] = 0.5$ and the integrals above are replaced by sums. Hence if $\beta_1^1<0$, the theoretical PIP becomes
\begin{align*}
&\probf{Y^*|X^*}{\frac{m^1(\mb\beta^1)+ m^0(\mb\beta^0)}{2} < Y^*|X^*=0}P(X^* = 0) + \\
&\probf{Y^*|X^*}{\frac{m^1(\mb\beta^1)+ m^0(\mb\beta^0)}{2} > Y^*|X^*=1}P(X^* = 1),
\end{align*}
or in terms of the conditional distribution function:
\begin{equation*}
  0.5\left[1-F^*_{Y^*|X^* =0}\left(\frac{m^0(\mb\beta^0)+m^1(\mb\beta^1)}{2}\right)\right] + 0.5F^*_{Y^*|X^* =1}\left(\frac{m^0(\mb\beta^0)+m^1(\mb\beta^1)}{2}\right).
\end{equation*}
Substituting the unknown model parameters with their sample estimators results into equation \eqref{cond_pip_general}. In total similarity, if $\beta_1^1>0$:
\begin{equation*}
  0.5\ F^*_{Y^*|X^* =0}\left(\frac{m^0(\mb\beta^0)+m^1(\mb\beta^1)}{2}\right) + 0.5\left[1-F^*_{Y^*|X^* =1}\left(\frac{m^0(\mb\beta^0)+m^1(\mb\beta^1)}{2}\right)\right].
\end{equation*}

\subsection{Continuous covariate}

This setting is closely related to the two-sample case, in the sense that the two base models $m^0$ and $m^1$ have the same functional form, but $X^*$ is now assumed to follow a continuous distribution instead of a Bernoulli distribution. In particular, $m^0(\mb\beta^0,{X^*}) = \beta_0^0$, $m^1(\mb\beta^1,{X^*}) = \beta_0^1 + \beta_1^1 X^*$,  $Y^*|X^* \sim \mathcal{N}(m^1(\mb\beta^1,{X^*}),\sigma^2)$ and $X^* \sim U([a,b])$ for constants $a<b$. Again dependent on the sign of $\beta_1^1$ and the location of $x^*$ relative to the expected value $E[X^*]$, the integral in \eqref{pip_th} consists of two parts ($X^*$ either above or below $K = E[X^*]$ ). With the additional assumption that $Y$ follows $m^1$,  the theoretical PIP can be written as (with $f_{X^*} = \frac{1}{b-a}$):

\begin{align}
\label{theoretical_pip_simple_regression}
    p_\text{th}(\mb\beta^0,\mb\beta^1,F^*,\mb\nu) = \begin{cases}  
    \int_{a}^{K} \Phi\left(\frac{m^0-m^1}{2\sigma}\right) f_{X^*}(x^*) dx^* + \int_{K}^{b} \left[1 - \Phi\left(\frac{m^0-m^1}{2\sigma}\right)\right] f_{X^*}(x^*) dx^* &  \text{ if } \beta^1_1 > 0, \\
    \int_{a}^{K} \left[1 - \Phi\left(\frac{m^0-m^1}{2\sigma}\right)\right] f_{X^*}(x^*) dx^* + \int_{K}^{b} \Phi\left(\frac{m^0-m^1}{2\sigma}\right) f_{X^*}(x^*) dx^* &  \text{ if } \beta^1_1 < 0, \\
    0.5 &  \text{ if } \beta^1_1 = 0.
  \end{cases}
\end{align}

For the corresponding plug-in estimator $m^0(\hat{\mb\beta^0})-m^1(\hat{\mb\beta^1}) = \hat{\beta^1_1} (\bar{x}-x^*)$ and hence
\begin{align}
\label{plugin_pip_simple_regression}
    p_\text{th}(\hat{\mb\beta^0},\hat{\mb\beta^1},F^*) = \begin{cases} 
    \int_{a}^{K} \Phi\left(\frac{\beta^1_1 (\bar{x}-x^*)}{2\sigma}\right) f_{X^*}(x^*) dx^* + \int_{K}^{b} \left[1 - \Phi\left(\frac{\beta^1_1 (\bar{x}-x^*) }{2\sigma}\right)\right] f_{X^*}(x^*) dx^* &  \text{ if } \hat{\beta^1_1} > 0, \\
  \int_{a}^{K} \left[1 - \Phi\left(\frac{\beta^1_1 (\bar{x}-x^*) }{2\sigma}\right)\right] f_{X^*}(x^*) dx^* + \int_{K}^{b} \Phi\left(\frac{\beta^1_1 (\bar{x}-x^*) }{2\sigma}\right) f_{X^*}(x^*) dx^* & \text{ if } \hat{\beta^1_1} < 0. \\
  \end{cases}
\end{align}
Now, since $\hat{K} = \bar{X}$, an expression similar to \eqref{pip_C1} is obtained:
\begin{align}
\label{estimator_simple_regression}
    p_\text{th}(\hat{\mb\beta^0},\hat{\mb\beta^1},F^*) =   
    \int_{a}^{K} \Phi\left(\frac{\lvert \hat{\beta^1_1} (\bar{x}-x^*)\rvert}{2\sigma}\right) f_{X^*}(x^*) dx^* + \int_{K}^{b}  \Phi\left(\frac{\lvert \hat{\beta^1_1} (\bar{x}-x^*) \rvert}{2\sigma}\right) f_{X^*}(x^*) dx^* , 
\end{align}
which can be linked to the two-sided p-value for testing whether $\beta_1^1 = 0$, using the fact that $\lvert \hat{\beta^1_1}\rvert = se(\hat{\beta^1_1})\Phi^{-1}(1-0.5 p_p)$.

\subsection{Relation with the conditional PIP}

In the paper, it is observed that the plug-in estimator of the theoretical PIP coincides with the conditional PIP C1. In the case of a normal linear regression model, C1 was found to correspond to
\[
  p_{\text{cond};1}(\hat{\mb\beta^0},\hat{\mb\beta^1},\hat{F}^*,\hat{\mb\nu}) = \Phi\left(\frac{\lvert \hat{\beta_1^1}\rvert}{4\hat{\sigma}_1}\right),
\]
Upon using a Taylor series expansion (with $\sigma_1$ assumed to be known) and assuming that $\hat{\beta_1^1}$ is a consistent estimator of $\beta_1^1$, with $\beta_1^1 \neq 0$,
\begin{eqnarray*}
  p_{\text{cond};1}(\hat{\mb\beta^0},\hat{\mb\beta^1},\hat{F}^*,\hat{\mb\nu})
    &=& \Phi\left(\frac{\lvert {\beta_1^1}\rvert}{4\sigma_1}\right)
        + (\hat\beta_1^1 - \beta_1^1) \frac{1}{4\sigma_1} \phi\left(\frac{\lvert 
 {\beta_1^1}\rvert}{4\sigma_1}\right) \frac{\beta_1^1}{\lvert \beta_1^1\rvert} + o_P(n^{-1/2}).
\end{eqnarray*}

Hence, as $n\rightarrow \infty$,
\[
  p_{\text{cond};1}(\hat{\mb\beta^0},\hat{\mb\beta^1},\hat{F}^*,\hat{\mb\nu}) \convProb \Phi\left(\frac{\lvert\beta_1^1\rvert}{4\sigma_1}\right) = p_\text{th}(\mb\beta^0,\mb\beta^1,F^*,\mb{\nu}),
\]
and so the plug-in PIP estimator is consistent as well. 

As an additional result, we see that 
\begin{eqnarray*}
\sqrt{n}\left( p_{\text{cond};1}(\hat{\mb\beta^0},\hat{\mb\beta^1},\hat{F}^*,\hat{\mb\nu})-p_\text{th}(\mb\beta^0,\mb\beta^1,F^*,\mb{\nu}) \right) &=& \sqrt{n}\left( p_{\text{cond};1}(\hat{\mb\beta^0},\hat{\mb\beta^1},\hat{F}^*,\hat{\mb\nu})-\Phi\left(\frac{\lvert\beta_1^1\rvert}{4\sigma_1}\right) \right) \\
&=& \sqrt{n}(\hat\beta_1^1 - \beta_1^1) \frac{1}{4\sigma_1} \phi\left(\frac{\lvert 
 {\beta_1^1}\rvert}{4\sigma_1}\right) \frac{\beta_1^1}{\lvert \beta_1^1\rvert} + \sqrt{n} o_P(n^{-1/2})
\end{eqnarray*}

In the latter expression, $\sqrt{n} o_P(n^{-1/2})$ approaches 0 as $n\rightarrow \infty$, while $\sqrt{n}(\hat\beta_1^1 - \beta_1^1)$ approaches a well-determined limiting normal distribution $Z$ with mean 0 and variance determined by the Fisher information. The term  $\frac{\beta_1^1}{\lvert \beta_1^1\rvert}$ equals -1 or 1 and hence does not alter the variance and leaves the mean at 0. This shows that, as $n\rightarrow \infty$,
\[
   \sqrt{n}\left( p_{\text{cond};1}(\hat{\mb\beta^0},\hat{\mb\beta^1},\hat{F}^*,\hat{\mb\nu})-p_\text{th}(\mb\beta^0,\mb\beta^1,F^*,\mb{\nu}) \right) \convDistr Z \frac{1}{4\sigma_1} \phi\left(\frac{\lvert \beta_1^1 \rvert}{4\sigma_1}\right),
\]
where $Z$ is the limiting normal distribution of $n^{1/2}(\hat{\beta_1^1} - \beta_1^1)$.

\section{Derivations related to the expected PIP}
\label{A_expected}

For the ordinary linear regression model of the shape $Y = X\beta + \epsilon$, with $\epsilon \sim \mathcal{N}(0,\sigma^2)$ the variance covariance matrix of the estimated coefficients is given by $\sigma^2(X^TX)^{-1}$. 

\subsection{Binary covariate}
For the case study in Section \ref{Example_TwoSample}, this leads to
\[
\sigma^2(X^TX)^{-1}=  \frac{\sigma^2}{n*\sum_i x_i^2 - (\sum_i x_i)^2}
\begin{pmatrix}
\sum_i x_i^2 & -\sum_i x_i \\
-\sum_i x_i & n
\end{pmatrix} = 
\sigma^2 \begin{pmatrix}
\frac{2}{n} & -\frac{2}{n} \\
-\frac{2}{n} & \frac{4}{n}
\end{pmatrix}.
\]
Moreover, as both models $m^0$ and $m^1$ are fitted on the same data, it is observed that
\[
\hat{\beta_0^0} = \bar{Y} = \hat{\beta_0^1} +\hat{\beta_1^1}\bar{X} = \hat{\beta_0^1} +\hat{\beta_1^1}\probh{X^* = 1}.
\]
With $E_1 = m^1(\hat{\mb\beta}^1)-Y^*$ and $E_0 = m^0(\hat{\mb\beta}^0)-Y^*$, it is then shown that the expected value and variance of $E_1$ are given by
\[
\Ef{Y^*,\cal{O}}{E_1 \mid X^*} = \Ef{Y^*,\cal{O}}{m^1(\hat{\mb\beta^1}) - Y^* \mid X^*} = \beta_0^1 + \beta_1^1 X^* - \beta_0^1 - \beta_1^1 X^*=0   
\]
and
\begin{align*}
\var{E_1} = \var{m^1(\hat{\mb\beta^1}) - Y^* \mid X^*} &= \var{\hat{\beta_0^1}}+ {X^*}^{2}\var{\hat{\beta_1^1}} + \var{Y^* \mid X^*} +2{X^*}\cov{\hat{\beta_0^1},\hat{\beta_1^1}}\\ &= \sigma^2(2/n+x^*\frac{4}{n} -2x^*\frac{2}{n}+1)  \\ &=  \sigma^2(2/n+1).
\end{align*}
Similarly, for $E_0$ it is obtained that
\begin{align*}
\Ef{Y^*,\cal{O}}{E_0 \mid X^*} &= \Ef{Y^*,\cal{O}}{m^0(\hat{\mb\beta^0}) - Y^*} = \beta_0^1 + \beta_1^1 \bar{X} - \beta_0^1 - \beta_1^1 X^*\\
&= \begin{cases}  
    \beta_1^1 \bar{X} & \text{if} X^* = 0 \\
    -\beta_1^1 \bar{X} & \text{if} X^* = 1 .
  \end{cases}
\end{align*}
and
\begin{align*}
\var{E_0 \mid X^*} &=  \var{m^0(\hat{\mb\beta^0}) - Y^* \mid X^*}\\
&= \var{\hat{\beta_0^0}} + \var{Y^*\mid X^*} \\ 
&= \var{\hat{\beta_0^1} + 0.5 \hat{\beta_1^1}} + \var{Y^*\mid X^*}  \\
&=  \var{\hat{\beta_0^1}} + 0.25\var{\hat{\beta_1^1}} + \Cov{\hat{\beta_0^1},\hat{\beta_1^1}} + \var{Y^*\mid X^*} \\ 
&= \sigma^2(1/n+1). 
\end{align*}
Finally, also the covariance between $E_1$ and $E_0$ is obtained as
\begin{align*}
\Cov{E_1,E_0 \mid X^*} &=  \Cov{\hat{\beta_0^1} + \hat{\beta_1^1}X^* - Y^*,\hat{\beta_0^1} + 0.5\hat{\beta_1^1} - Y^* \mid X^*}\\
&= \var{\hat{\beta_0^1}} +X^*\Cov{\hat{\beta_0^1},\hat{\beta_1^1}}+0.5\Cov{\hat{\beta_0^1},\hat{\beta_1^1}}+0.5X^*\var{\hat{\beta_1^1}} + \var{Y^*} \\
&= \begin{cases}  
    \sigma^2(1+\frac{1}{n}) & \text{if } X^* = 0 \\
    \sigma^2(1+\frac{1}{n})  & \text{if } X^* = 1 .
  \end{cases}
\end{align*}
Together, this results into expression \eqref{expected_simple_reg}.

\subsection{Continuous covariate}
In full analogy for a continuous covariate $X$, the least squares regression coefficient estimators are normally distributed:
\begin{equation}
\begin{pmatrix} 
\hat{\beta_0^1}  \\
\hat{\beta_1^1}
\end{pmatrix} \sim \mathcal{N}\left(
\begin{pmatrix} 
\beta_0^1  \\
\beta_1^1
\end{pmatrix},\sigma^2 \begin{pmatrix}
\frac{\sum_i x_i^2}{n\sum_i x_i^2 - (\sum_i x_i)^2} & \frac{-\sum_i x_i}{n\sum_i x_i^2 - (\sum_i x_i)^2} \\
\frac{-\sum_i x_i}{n\sum_i x_i^2 - (\sum_i x_i)^2} & \frac{n}{n\sum_i x^2_i - (\sum_i x_i)^2}
\end{pmatrix}\right).
\end{equation}
Hence the standard error of $\hat{\beta^1_1}$ is given by $\sqrt{\frac{n\sigma^2}{n\sum_i x^2_i - (\sum_i x_i)^2}}$ and expression \eqref{estimator_simple_regression} is further reduced to
\begin{align}
\label{estimator_simple_regression_reduced}
    p_\text{th}(\hat{\mb\beta^0},\hat{\mb\beta^1},F^*) =   
    \int_{a}^{b} \Phi\left(\frac{\Phi^{-1}(1-0.5 p_p) \lvert\bar{x}-x^* \rvert}{2\sqrt{\sum_i x_i^2 - n\bar{x}^2}}\right) f_{X^*}(x^*) dx^*.
\end{align}

For the expected PIP, the expression is similar to \eqref{expected_simple_reg}, where $X^*$ is now considered to be uniformly distributed on the $[a,b]$ interval and the mean of $E_2$ now depends on $X^*$:
\begin{equation}
\label{expected_simple_reg_full}
E = \begin{pmatrix} 
E_1  \\
E_2
\end{pmatrix}\sim \int_a^b\mathcal{N}\left(
\begin{pmatrix} 
0  \\
\beta_1^1(x^*-\bar{x})
\end{pmatrix},\Sigma_{E_1,E_2} \right) f_{X*}(x^*)  ,
\end{equation}
where 
\[
\Sigma_{E_1,E_2}  = \begin{pmatrix}
        \sigma^2+  \sigma^2_{\hat{\beta_0^1}}+ {x^*}^2\sigma^2_{\hat{\beta_1^1}}+2x^*\sigma^2_{\hat{\beta_0^1},\hat{\beta_1^1}}
         &  \sigma^2+\sigma^2_{\hat{\beta_0^1}}+ x^*\sigma^2_{\hat{\beta_0^1},\hat{\beta_1^1}}+\bar{x}*\sigma^2_{\hat{\beta_0^1},\hat{\beta_1^1}}+\bar{x}x^*\sigma^2_{\hat{\beta_1^1}} \\
 \sigma^2+\sigma^2_{\hat{\beta_0^1}}+ x^*\sigma^2_{\hat{\beta_0^1},\hat{\beta_1^1}}+\bar{x}\sigma^2_{\hat{\beta_0^1},\hat{\beta_1^1}}+\bar{x}*x^*\sigma^2_{\hat{\beta_1^1}}& \sigma^2 +  \sigma^2_{\hat{\beta_0^0}} 
        
\end{pmatrix}
\]
and 
\[
\sigma^2_{\hat{\beta_0^0}}  = \text{Var}(\hat{\beta_0^1}+\hat{\beta_1^1}\bar{X}) = \sigma^2_{\hat{\beta_0^1}} + \bar{X}^2 \sigma^2_{\hat{\beta_1^1}} +2\bar{X} \sigma^2_{\hat{\beta_0^1},\hat{\beta_1^1}}.
\]

\section{Relationships between the PIP and other statistics}
\label{A_relations}

In this appendix, details are provided about the relationship of the PIP with other statistics for comparing models.

\subsection{P-values}

The asymptotic result of Equation (\ref{eq:PAssymptotic}) follows from \cite{Lambert1982}, where lemma 4.2 states (with $p_n$ the $p$-value based on a sample of size $n = n_1 + n_2$), as $n \rightarrow\infty`$,
\[
  n^{-1/2} \left(\ln p_n + nc \right) \convDistr N(0,\tau^2),
\]
with 
\begin{eqnarray*}
  c &=& \frac{1}{2} \ln\left( 1+\lambda(1-\lambda) \Delta^2\right) \\
  \lambda &=& \lim_{n\rightarrow\infty} \frac{n_1}{n} \\
  \tau^2 &=& \frac{\lambda(1-\lambda)\Delta^2}{1+\lambda(1-\lambda)\Delta^2} \\
  \Delta &=& \frac{\beta}{\sigma}.
\end{eqnarray*}

This results in,
\[
  n^{-1} \ln p_n \convProb c=-\frac{1}{2}\ln\left(1+\frac{1}{4}\Delta^2\right) = -\frac{1}{2}\ln\left(1+\frac{1}{4}\frac{\beta^2}{\sigma^2}\right).
\]

Combining the results for the convergence of the plug-in PIP estimator and the large $n$ results for the $p$-value, we can write 
\[
  n^{-1} \ln p_n \convProb -\frac{1}{2}\ln\left(1+4\left(\Phi^{-1}\left(\text{PIP}\right)\right)^2\right).
\]

\subsection{Difference in MSE between full and null model}

Also for MSE difference between the estimated full model and null model, an exact relation with the conditional PIP can be derived. For model $m^1$, it is known that:

\begin{align*}
MSE_1 &= \Ef{X^*,Y^*}{\left(Y^* -  (\hat{\beta_0^1}+\hat{\beta_1^1x^*})\right)^2 \mid {\cal{O}}} \\
&= 0.5\Ef{Y^*,X^*=0}{\left(Y^* - \hat{\beta_0^1} \right)^2 \mid {\cal{O}}}  + 0.5\Ef{Y^*,X^*=1}{\left(Y^* -  (\hat{\beta_0^1}+\hat{\beta_1^1}) \right)^2 \mid {\cal{O}}}\\
&= 0.5\Ef{Y^*,X^*=0}{{Y^*}^2 -2\hat{\beta_0^1}Y^*+\hat{\beta_0^1}^2} + 0.5\Ef{Y^*,X^*=1}{{Y^*}^2 -2\hat{\beta_0^1}Y^*-2\hat{\beta_1^1}Y^*+\hat{\beta_0^1}^2+2\hat{\beta_0^1}\hat{\beta_1^1}+\hat{\beta_1^1}^2} \\
&= 0.5\Ef{Y^*,X^*=0}{{Y^*}^2} - \hat{\beta_0^1}\Ef{Y^*,X^*=0}{Y^*} + 0.5\hat{\beta_0^1}^2 + 0.5\Ef{Y^*,X^*=1}{{Y^*}^2}-\hat{\beta_0^1}\Ef{Y^*,X^*=1}{Y^*}\\
& \ \ \  -\hat{\beta_1^1}\Ef{Y^*,X^*=1}{Y^*}+ 0.5\hat{\beta_0^1}^2+ 0.5\hat{\beta_1^1}^2+ \hat{\beta_0^1}\hat{\beta_1^1}
\end{align*}

and similarly for model $m^0$, it is seen that

\begin{align*}
MSE_0 &= \Ef{X^*,Y^*}{\left(Y^* -  \hat{\beta_0^0}\right)^2 \mid {\cal{O}}} \\
&= 0.5\Ef{Y^*,X^*=0}{\left(Y^* -  \hat{\beta_0^0}\right)^2 \mid {\cal{O}}}  + 0.5\Ef{Y^*,X^*=1}{\left(Y^* -  \hat{\beta_0^0}\right)^2 \mid {\cal{O}}}\\
&= 0.5\Ef{Y^*,X^*=0}{{Y^*}^2} - \hat{\beta_0^0}\Ef{Y^*,X^*=0}{Y^*} + 0.5\hat{\beta_0^0}^2 \\
& \ \ \ + 0.5\Ef{Y^*,X^*=1}{{Y^*}^2} - \hat{\beta_0^0}\Ef{Y^*,X^*=1}{Y^*} + 0.5\hat{\beta_0^0}^2 \\
\end{align*}

Subtracting $MSE_0$ from $MSE_1$, we obtain
\begin{align*}
\Delta MSE &= MSE_1 - MSE_0 \\
&= \frac{\hat{\beta_1^1}^2}{4} - \frac{\hat{\beta_1^1}\beta_1^1}{2} \convProb \frac{ - {\beta_1^1}^2}{4}
\label{True_MSE_diff_PIP_relation}
\end{align*}
which can be estimated by $\widehat{\Delta MSE} =  \frac{-\hat{\beta_1^1}^2}{4} = \frac{- \lvert \hat{\beta_1^1}\rvert ^2}{4}$.

Upon using that $\widehat{PIP} = \Phi(\frac{\lvert \hat{\beta_1^1}\rvert}{4\sigma})$, we obtain the relation between $\widehat{\Delta MSE}$ and $\widehat{PIP}$ as:
\[
\widehat{\Delta MSE} = -4*\sigma^2*\Phi^{-1}(\widehat{PIP})^2.
\]

\subsection{Measure of overlap between predictive distributions}

For the two-sample case with a known variance $\sigma_1^2$, following \cite{Clarke2012}, the predictive distribution for a prediction of a new observation based on model $m^1$ is given by a normal distribution with mean $\hat{\beta_0^1} + \hat{\beta_1^1}x^* $ and variance $\sigma_1^2(1+\frac{2}{n})$. This corresponds to the distribution of $E_1$ in equation \ref{expected_simple_reg} related to the Expected PIP. According to \cite{Billheimer2019}, the importance of the explanatory variable can be summarized by graphing this predictive distribution at key values of $X$, which in the current case corresponds to $x^* = 0$ (leading to $\mu_1 = \hat{\beta_0^1}$) and $x^* = 1$ (leading to $\mu_2 = \hat{\beta_0^1}+ \hat{\beta_1^1}$). The less overlap between these distributions, the more important $X$. Following the definition by \cite{Rom1996}, the overlap between two Gaussian densities with means $\mu_1$ and $\mu_2$ and (common) variance $\sigma_1^2(1+\frac{2}{n})$ is expressed by:
\begin{align*}
Ovl &= 1-\Phi(\frac{\frac{\mu_1+\mu_2}{2}-\min(\mu_1,\mu_2)}{\sigma_1\sqrt{1+\frac{2}{n}}})+\Phi(\frac{\frac{\mu_1+\mu_2}{2}-\max(\mu_1,\mu_2)}{\sigma_1\sqrt{1+\frac{2}{n}}}) \\
 &=\begin{cases}  2\Phi(\frac{-\hat{\beta_1^1}}{2\sigma_1}\sqrt{1+\frac{2}{n}}) &  \text{when } \hat{\beta_1^1} > 0, \\
   2\Phi(\frac{\hat{\beta_1^1}}{2\sigma_1\sqrt{1+\frac{2}{n}}}) &  \text{when } \hat{\beta_1^1} < 0.
  \end{cases}
\end{align*}
With $p_{\text{th};1}(\hat{\mb\beta^0},\hat{\mb\beta^1},\hat{F}^*,\mb\nu) = \Phi\left(\frac{\lvert \hat{\beta_1^1}\rvert}{4\sigma_1}\right)$ it is readily derived that 
\[
Ovl = 2\Phi\left(\frac{-2\Phi^{-1}(p_{\text{th};1}(\hat{\mb\beta^0},\hat{\mb\beta^1},\hat{F}^*,\mb\nu))}{\sqrt{1+\frac{2}{n}}}\right).
\]

\newpage
\section{Additional simulation results}
\label{C_Sims}

\subsection{Two-sample setting}

\begin{figure}[htbp]
    \centering
    \includegraphics[scale=0.18]{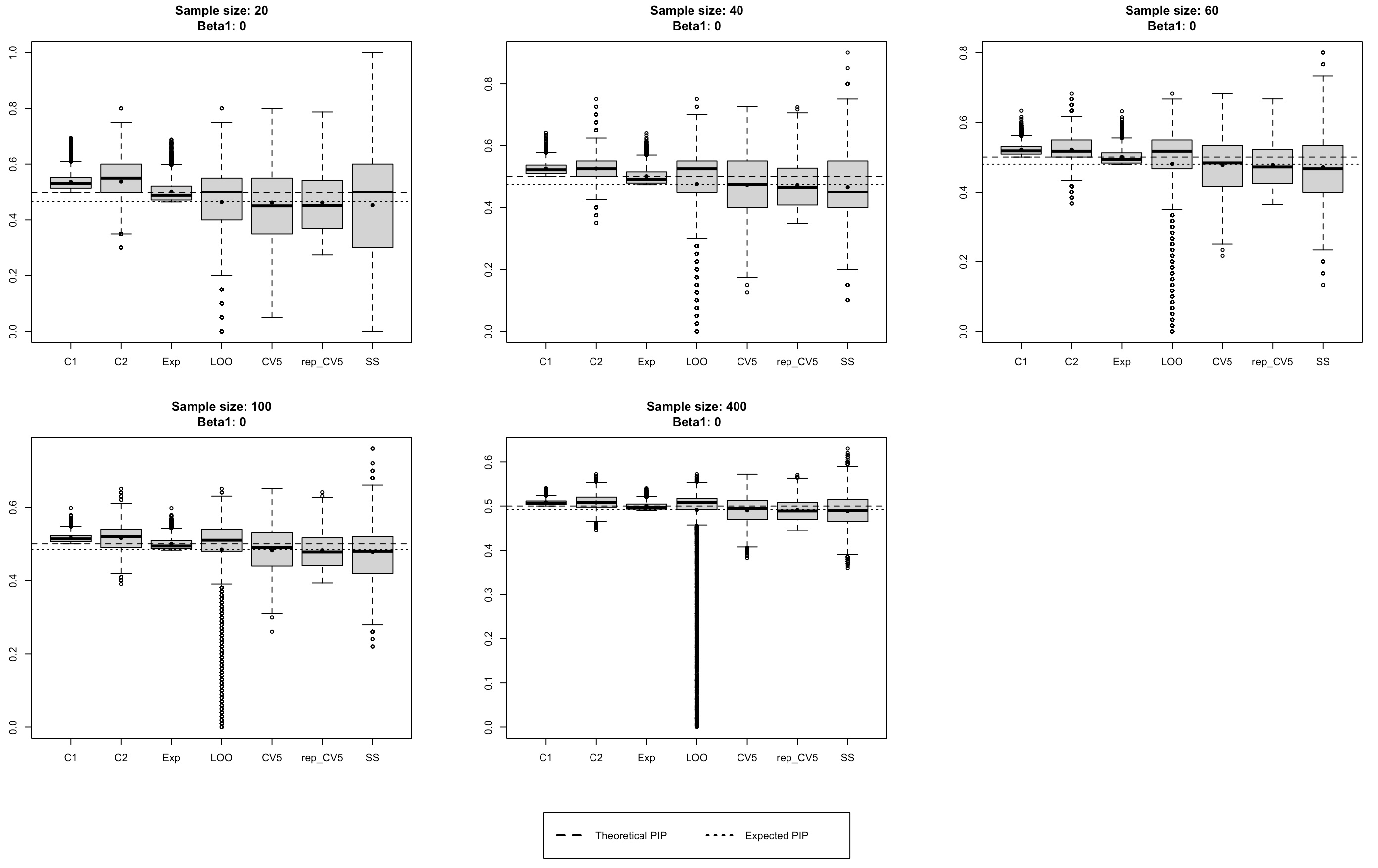}
    \caption{Simulation results for effect size $\beta_1^1=0$. The boxplots show the distributions of the various PIP estimates over 10000 simulation runs. $C1$ and $C2$ refer to $p_{\text{th};1}(\hat{\mb\beta^0},\hat{\mb\beta^1},\hat{F}^*,\hat{\mb\nu})$ and $p_{\text{th};2}(\hat{\mb\beta^0},\hat{\mb\beta^1},\hat{F}^*,\hat{\mb\nu})$, respectively, while Exp refers to the plug-in estimate for the expected PIP. LOO, CV5, rep$\_$CV5 and SS refer to estimates based on leave-one-out CV, 5-fold CV, repeated 5-fold cross-validation and split sampling, respectively. The true theoretical (dashed) and expected PIP (dotted) are indicated as horizontal reference lines. The mean over the simulation runs is indicated by the solid dot in the boxplot. Effect size = 0.}
    \label{Sim0_full}
\end{figure}

\begin{figure}[htbp]
    \centering
    \includegraphics[scale=0.18]{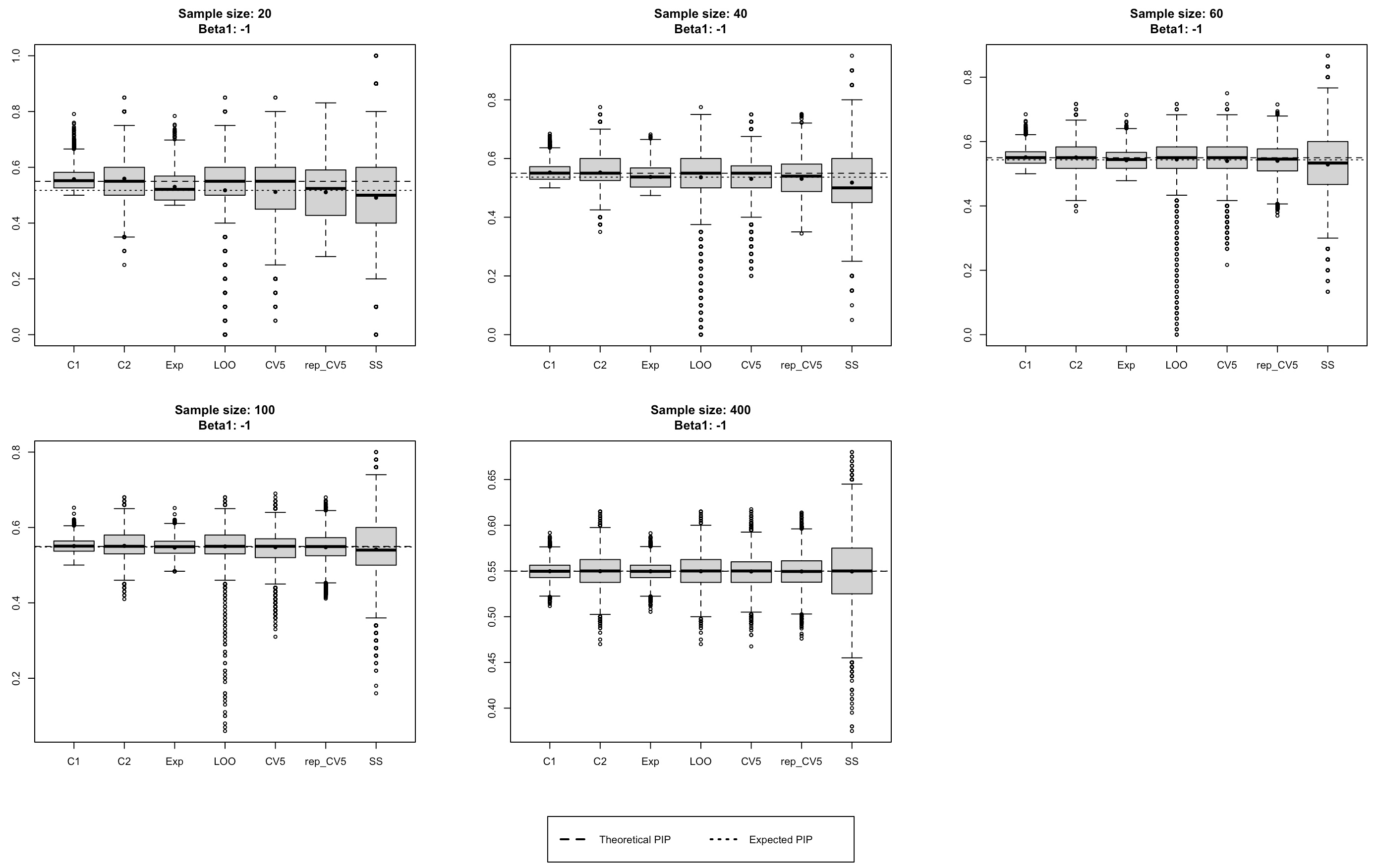}
    \caption{Simulation results for effect size $\beta_1^1=-1$. See the caption of Figure \ref{Sim0_full} for more details.}
    \label{Sim1_full}
\end{figure}

\begin{figure}[htbp]
    \centering
    \includegraphics[scale=0.18]{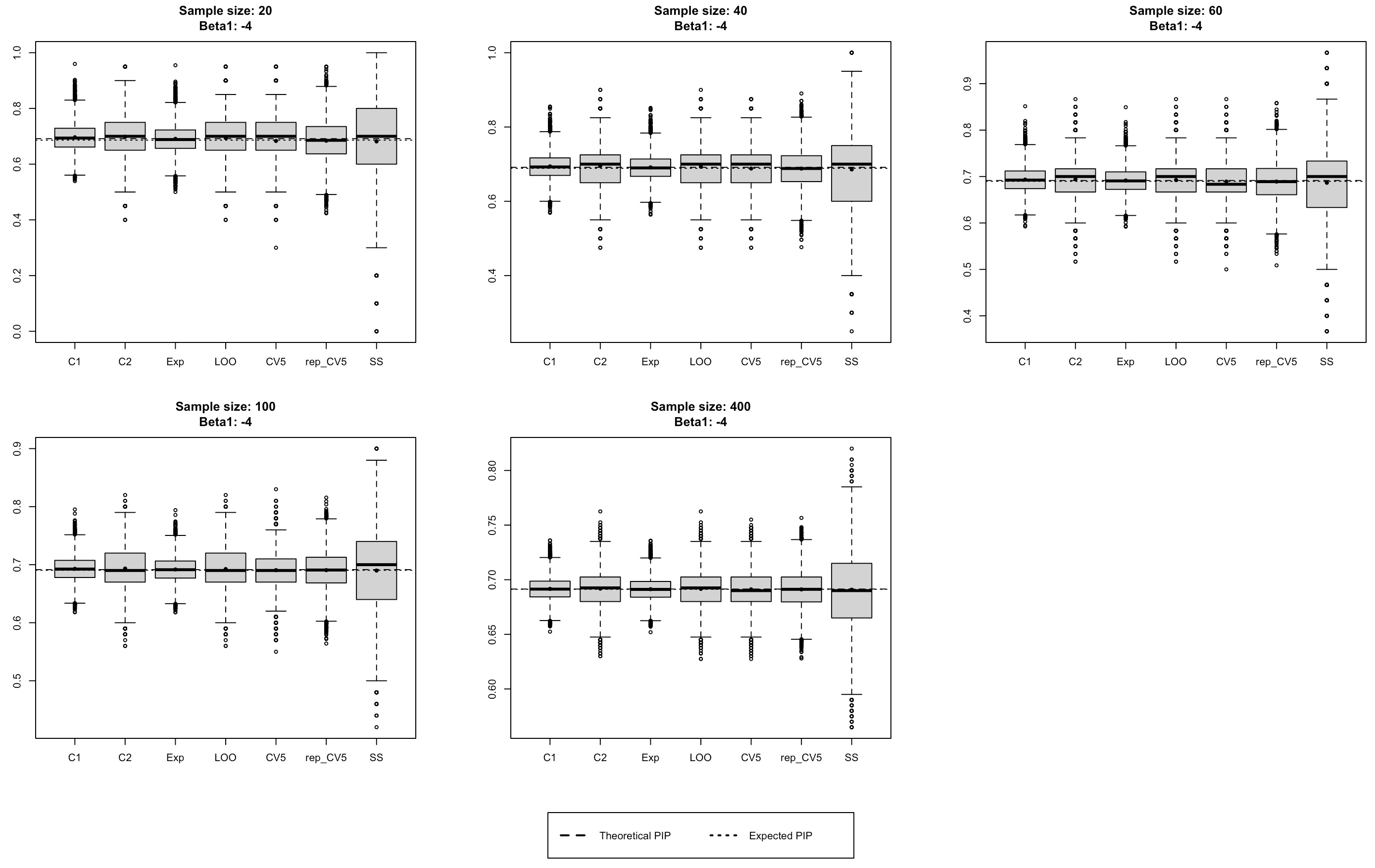}
    \caption{Simulation results for effect size $\beta_1^1=-4$. See the caption of Figure \ref{Sim0_full} for more details.}
    \label{Sim4_full}
\end{figure}

\begin{figure}[htbp!]
    \centering
    \includegraphics[scale=0.18]{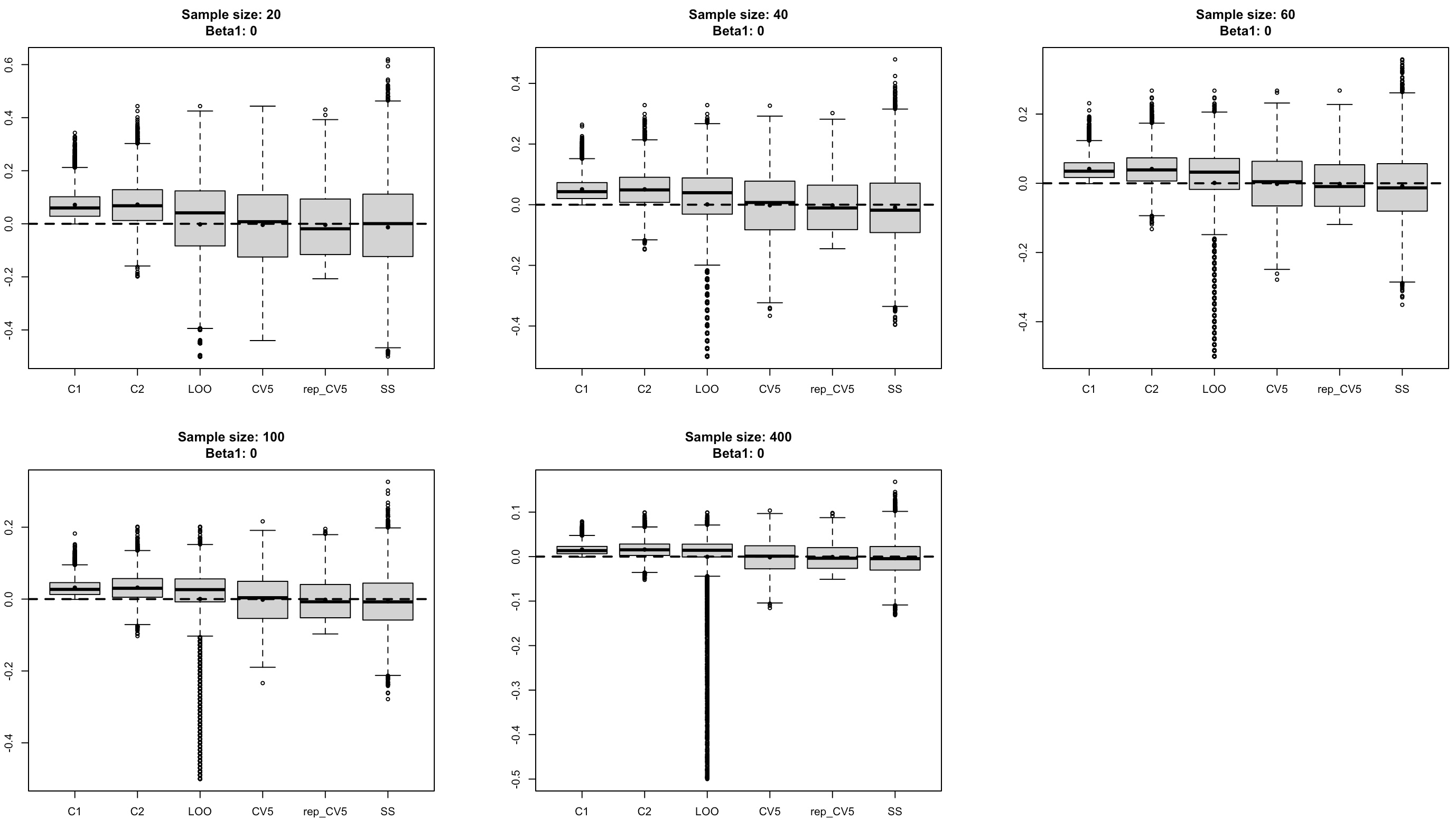}
    \caption{Comparing the difference between the empirical conditional PIP and the PIP estimators. Effect size = 0. Abbreviations as in Figure \ref{Sim0_full}.}
    \label{Sim0_comp_full}
\end{figure}

\begin{figure}[htbp!]
    \centering
    \includegraphics[scale=0.18]{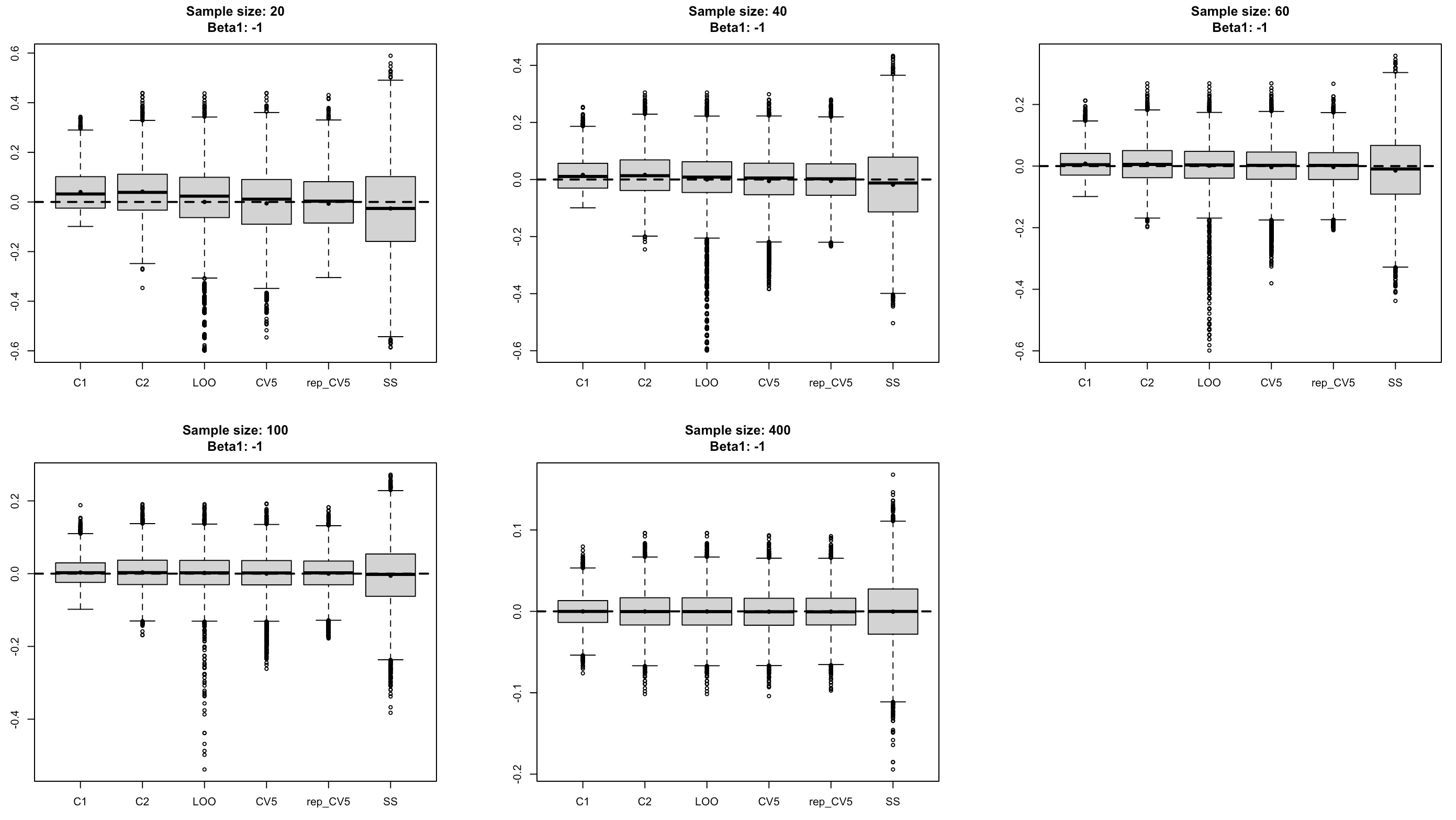}
    \caption{Comparing the difference between the empirical conditional PIP and the PIP estimators. Effect size = -1. Abbreviations as in Figure \ref{Sim0_full}.}
    \label{Sim1_comp_full}
\end{figure}

\begin{figure}[htbp!]
    \centering
    \includegraphics[scale=0.18]{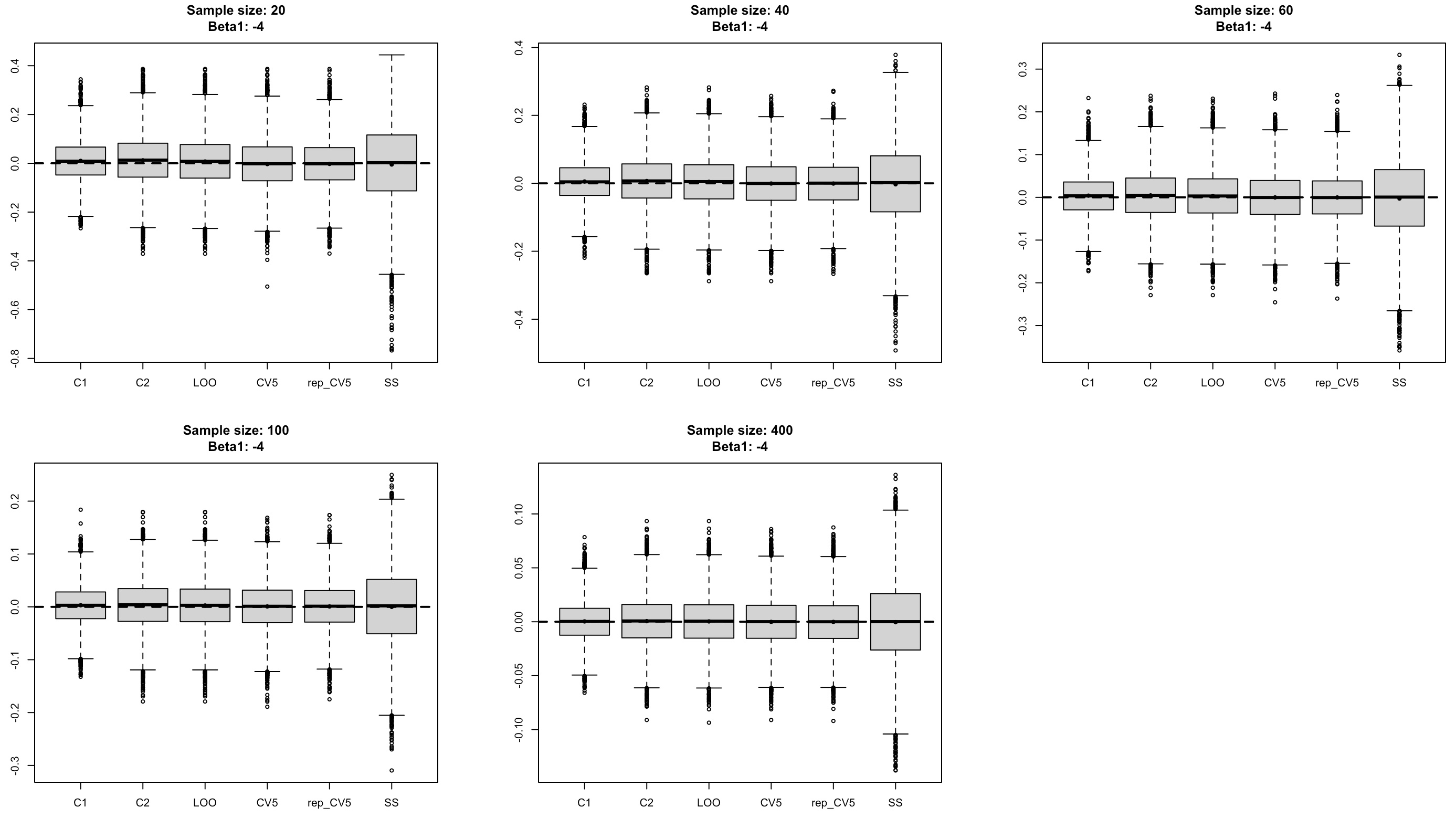}
    \caption{Comparing the difference between the empirical conditional PIP and the PIP estimators. Effect size = -4. Abbreviations as in Figure \ref{Sim0_full}.}
    \label{Sim4_comp_full}
\end{figure}

\begin{figure}[htbp]
    \centering
    \includegraphics[scale=0.5]{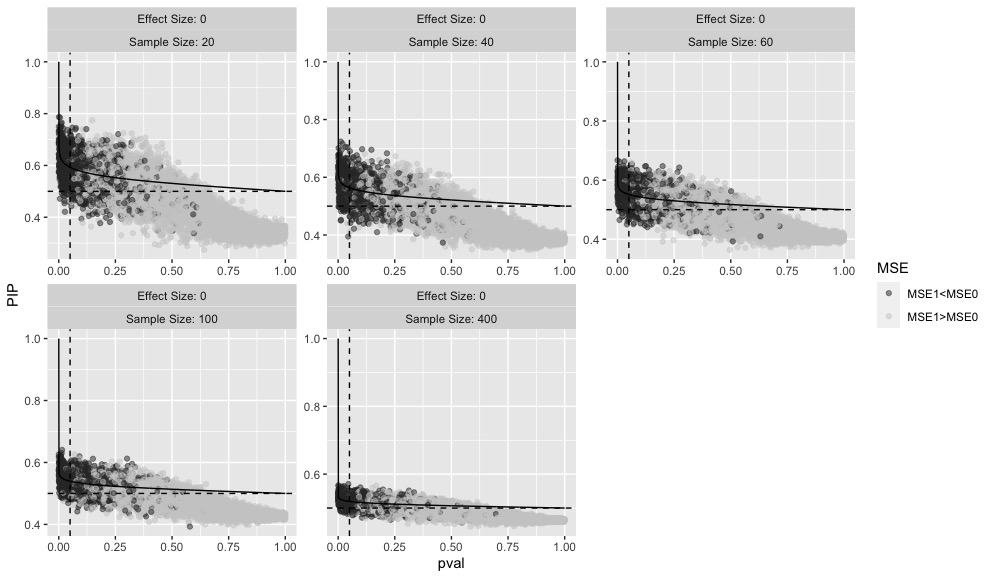}
    \caption{Relation between repeated 5-fold CV PIP and the p-value for testing $\beta_1^1 = 0$. Coloring based on the difference between MSE of $m^1$ and $m^0$. Solid line indicates exact relation obtained in \eqref{expr:estimator_reduced2}. Effect size = 0.}
    \label{Sim0_rel_full}
\end{figure}

\begin{figure}[htbp]
    \centering
    \includegraphics[scale=0.5]{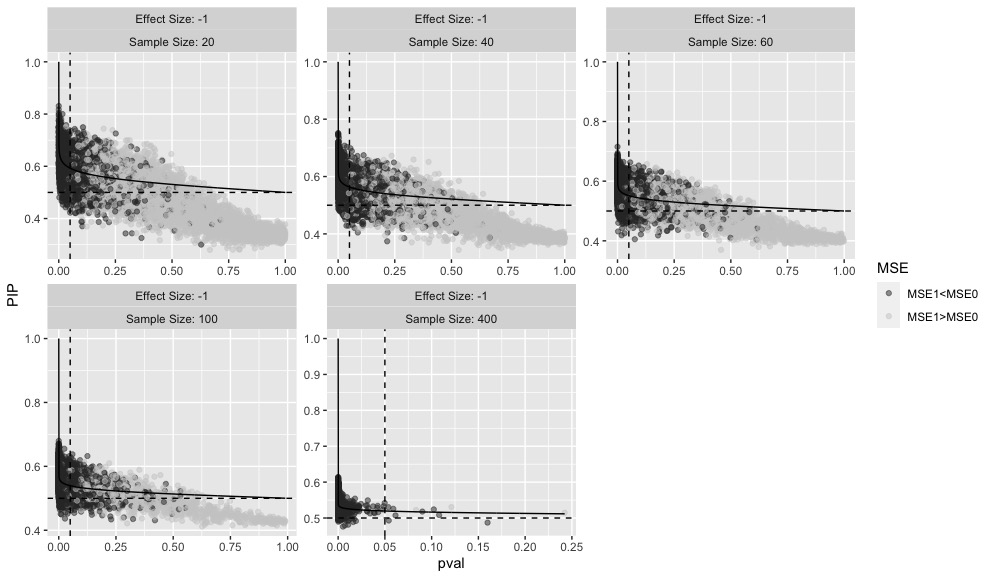}
    \caption{Relation between repeated 5-fold CV PIP and the p-value for testing $\beta_1^1 = 0$. Coloring based on the difference between MSE of $m^1$ and $m^0$. Solid line indicates the exact relation obtained in \eqref{expr:estimator_reduced2}. Effect size = -1.}
    \label{Sim1_rel_full}
\end{figure}

\begin{figure}[htbp]
    \centering
    \includegraphics[scale=0.5]{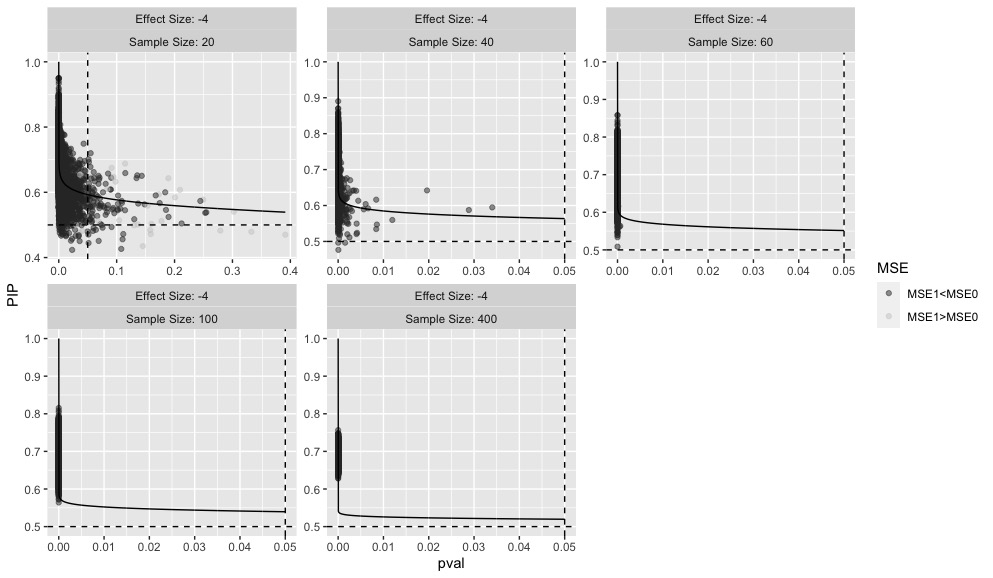}
    \caption{Relation between repeated 5-fold CV PIP and the p-value for testing $\beta_1^1 = 0$. Coloring based on the difference between MSE of $m^1$ and $m^0$. Solid line indicates the exact relation obtained in \eqref{expr:estimator_reduced2}. Effect size = -4.}
    \label{Sim4_rel_full}
\end{figure}
\newpage
\subsection{Nonlinear setting with gradient boosting machines}

In this setting, data are simulated from a nonlinear model
\[
y=\left|4x_1\right|^{3x_4}+5x_2+(2x_3)^{x_5} +\varepsilon, 
\]
where $x_1 \sim U([0,6])$ and $x_2 \sim \mathcal{N}(0;1)$ both rounded to the nearest integer, $x_3 \sim Bern(0.5)$, $x_4 \sim U([0,1])$ and $x_5 \sim U([1,2])$, rounded to the first decimal. The error $\varepsilon$ is sampled from a Gaussian distribution with mean zero and standard deviation 1.6. 

Two models are fitted:
\begin{itemize}
    \item $m^0$ = gradient boosting machine with variables $x_1$, $x_2$ and $x_3$
    \item $m^1$ = gradient boosting machine with all variables.
\end{itemize}
The hyperparameters are taken to be fixed with values for the interaction depth equal to 2, number of trees equal to 50, 
shrinkage equal to 0.1 and a minimum of 2 observations per node.

In the plot below, we compare with the empirical version of the conditional PIP, which is calculated from 1000000 draws of the true model above and consecutively calculating the mean proportion of times the estimated model $m^1$ provided better predictions than estimated model $m^0$.

\begin{figure}[htbp!]
    \centering
    \includegraphics[scale=0.30]{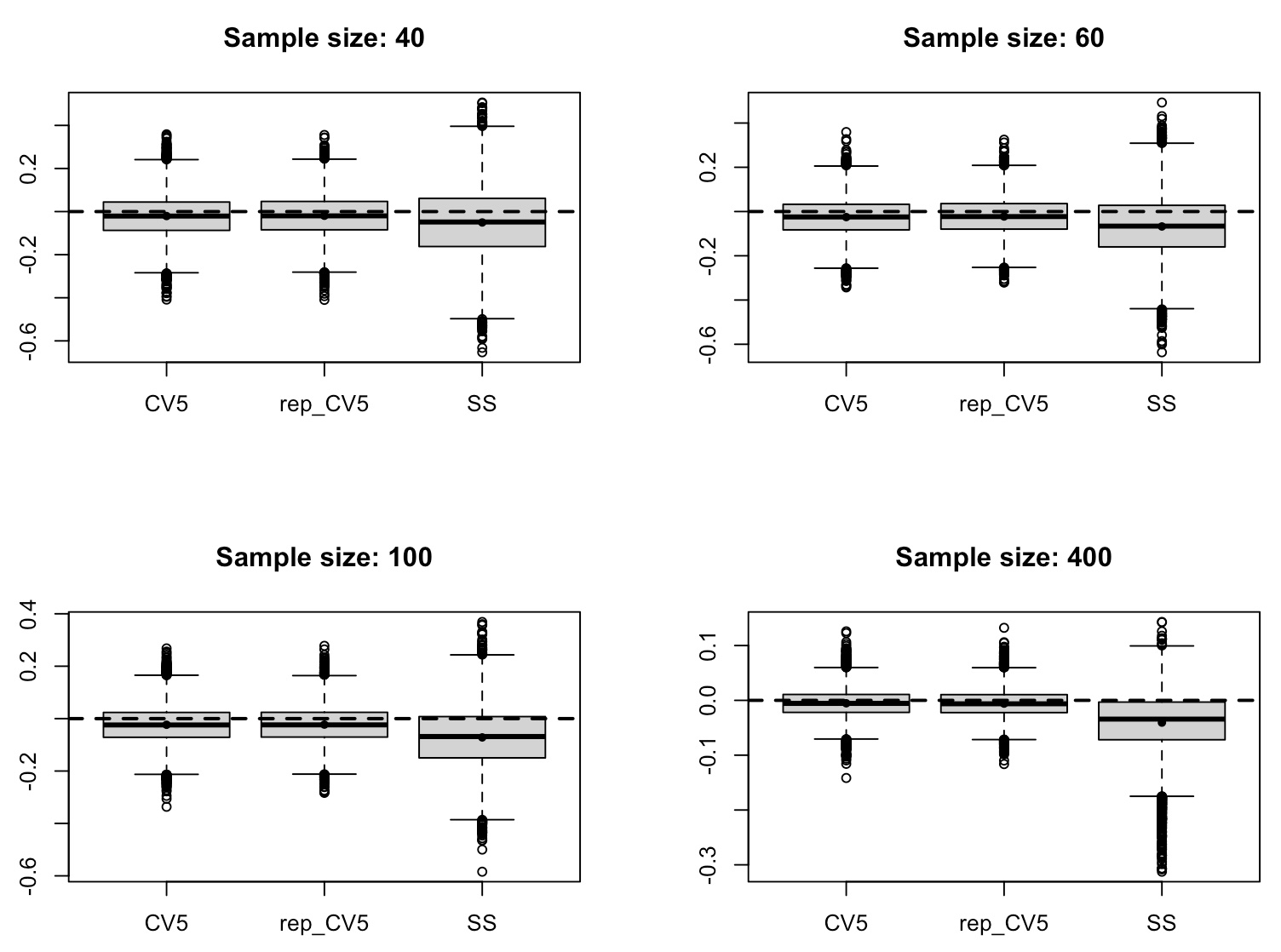}
    \caption{Comparing the difference between the empirical conditional PIP and the non-parametric PIP estimators based on two gradient boosting models.}
    \label{Sim_RF}
\end{figure}

From the earlier analysis, it is known that the CV estimators are targeted towards the expected PIP, which was slightly lower as compared to the conditional PIP. Again here, it is observed that both the CV approaches outperform the split sampling approach and are approximating the underlying conditional PIP as sample size increases. This shows the applicability of the non-parametric approaches, even in a more complex setting of machine learning models.

\newpage
\section{Settings for sampling application data}
\label{D_Application}
In the application section, four studies were considered that were part of a bigger discussion on replicability in \cite{Camerer2018}. The exact data could only be obtained for the study by \cite{Balafoutas2012}, that dealt with the comparison of proportions in two groups. In the remaining studies, dealing with the comparison of the mean of a continuous outcome in two groups, data were sampled based on the mentioned sample mean and standard deviation. A seed was chosen such that the resulting p-value was close to the originally reported one. Table \ref{Study_settings} provides more details on the nature of the data and used parameter settings.

\begin{table}[htbp]
\caption{Parameter settings for the four studies under consideration in the application of the PIP in the setting of study replicability based on \cite{Camerer2018}.}
\label{Study_settings}
\begin{center}
\footnotesize
\begin{tabular}{rrcccc}
Study & Outcome & Parameters & Seed& New p-value& Original p-value\\\hline
\cite{Gervais2012} &Gaussian&\makecell[l]{$n_1$ = 31 \ \ $\bar{x}_1$= 61.55 \ \  $s_1$ = 35.68 \\ $n_2$ = 26 \ \  $\bar{x}_2$= 41.42 \ \  $s_2$ = 31.47}&457 &0.0291& 0.029 \\
\cite{Ackerman2010} &Gaussian&\makecell[l]{$n_1$ = 26 \ \ $\bar{x}_1$= 5.80 \ \  $s_1$ = 0.76 \\ $n_2$ = 28 \ \  $\bar{x}_2$= 5.38 \ \  $s_2$ = 0.79}&62 &0.0488&0.049  \\
\cite{Balafoutas2012} &Binomial&\makecell[l]{$n_1$ = 36 \ \ $p_1= 30.6\%$ \\ $n_2$ = 36  \ \ $p_2= 58.3\%$} & - &0.0177&0.018 \\
\cite{Wilson2014} & Gaussian&\makecell[l]{$n_1$ = 15 \ \ $\bar{x}_1$= 3.20 \ \  $s_1$ = 2.23 \\ $n_2$ = 15 \ \  $\bar{x}_2$= 6.87 \ \  $s_2$ = 1.91}&62&$<$0.001 & $<$0.001 \\
\end{tabular}
\end{center}
\end{table}

\end{document}